\documentclass[prd,onecolumn]{revtex4}
\usepackage{dcolumn}
\usepackage{multirow}
\usepackage{graphicx}
\usepackage{amssymb}
\usepackage{bm}
\usepackage{hyperref}%for .pdf
\usepackage{epstopdf}%for .pdf
\usepackage{color}
\usepackage{mathrsfs}
\usepackage{amsmath,amssymb,amsthm}
\begin{document}
\title{Model-independent determination on $H_0$ using the latest $H(z)$ data}
\author{Deng Wang}
\email{Cstar@mail.nankai.edu.cn}
\affiliation{Theoretical Physics Division, Chern Institute of Mathematics, Nankai University,
Tianjin 300071, China}
\author{Xin-He Meng}
\email{xhm@nankai.edu.cn}
\affiliation{{Department of Physics, Nankai University, Tianjin 300071, China}}
\begin{abstract}
We perform the improved constraints on the Hubble constant $H_0$ by using the model-independent method, Gaussian Processes. Utilizing the latest 36 $H(z)$ measurements, we obtain $H_0=69.21\pm3.72$ km s$^{-1}$ Mpc$^{-1}$, which is consistent with the Planck 2015 and Riess et al. 2016 analysis at $1\sigma$ confidence level, and reduces the uncertainty from $6.5\%$ (Busti et al. 2014) to $5.4\%$. Different from the results of Busti et al. 2014 by only using 19 $H(z)$ measurements, our reconstruction results of $H(z)$ and the derived values of $H_0$ are independent of the choice of covariance functions.
\end{abstract}
\maketitle
\section{Introduction}
An urgent task in modern cosmology is to measure the Hubble constant $H_0$ accurately, since it brings substantially important information of the universe such as the size and age of the universe, the present cosmic expansion rate and the cosmic components. Due to the early determination by Hubble \cite{1}, the value of $H_0$ was believed to lie in the relatively large range [50, 100] km s$^{-1}$ Mpc$^{-1}$ for a long time \cite{2}. With the use of different calibration techniques and improved control of systematics, the first accurate value, $H_0=72\pm8$ km s$^{-1}$ Mpc$^{-1}$ \cite{3}, was given by the local measurements from the Hubble Space Telescope in 2001. After ten years, Riess et al. calibrated the Type Ia supernovae (SNe Ia) and obtained $H_0=73.8\pm2.4$ km s$^{-1}$ Mpc$^{-1}$ \cite{4} by using three indicators, i.e., the distance to NGC 4258 from a megamaser measurement, the trigonometric parallaxes measurements to the Milk Way (MW) Cepheids, Cepheid observations and a modified distance to the Large Magellanic Cloud (LMC). In 2012, there are three groups to measure the Hubble constant: Riess et al. got $H_0=75.4\pm2.9$ km s$^{-1}$ Mpc$^{-1}$ by utilizing the Cepheids in M31 \cite{5}; Freedman et al. obtained $H_0=74.3\pm2.1$ km s$^{-1}$ Mpc$^{-1}$ by using a mid-infrared calibration for the Cepheids \cite{6}; Ch\'{a}ves et al. got $H_0=74.3\pm3.1($random$)\pm2.9$ $($syst.$)$ km s$^{-1}$ Mpc$^{-1}$ by adopting HII regions and HII galaxies as distance indicators \cite{7}. Subsequently, in 2013, $H_0=67.3\pm1.2$ km s$^{-1}$ Mpc$^{-1}$ derived by Planck \cite{8} from the anisotropies of the cosmic microwave background (CMB) exhibits a strong tension with the local measurement from Riess et al. 2011 at 2.4$\sigma$ level. For the purpose to alleviate or resolve this tension, several groups implemented the measurements using different methods: Bennett et al. \cite{9} and Hinshaw et al. \cite{10} gave a $3\%$ determination, namely, $H_0=70.0\pm2.2$ km s$^{-1}$ Mpc$^{-1}$ by utilizing the nine-year Wilkinson Microwave Anisotropy Probe (WMAP9) data; Spergel et al. \cite{11} found $H_0=68.0\pm1.1$ km s$^{-1}$ Mpc$^{-1}$ by removing the $217\times217$ GHz detector set spectrum used in the Planck analysis; Fiorentino et al. \cite{12} obtained $H_0=76.0\pm1.9$ km s$^{-1}$ Mpc$^{-1}$ by using 8 new classical Cepheids observed in galaxies hosting SNe Ia; Different from the calibration method exhibited by Riess et al. 2011, Tammann et al. \cite{13} got a lower value $H_0=63.7\pm2.3$ km s$^{-1}$ Mpc$^{-1}$ by calibrating the SN Ia with the tip of red-giant branch (TRGB); Efstathiou \cite{14} obtained $H_0=70.6\pm3.3$ km s$^{-1}$ Mpc$^{-1}$ by revising the geometric maser distance to NGC 4258 from Humphreys et al. \cite{15} and using this indicator to calibrate the Riess et al. 2011 data; Rigault et al. \cite{16} gave $H_0=70.6\pm2.6$ km s$^{-1}$ Mpc$^{-1}$ by considering predominately star-forming environments.

The mid-redshift data can also act as an effective and complementary tool to determine the value of $H_0$. Combining them with the high-redshift CMB data, the uncertainties of $H_0$ can be reduced significantly. Through making full use of baryon acoustic oscillations (BAO) data, Cheng et al. \cite{17} found $H_0=68.0\pm1.1$ km s$^{-1}$ Mpc$^{-1}$ for the $\Lambda$CDM model. Utilizing the CMB and BAO data and assuming the six-parameter $\Lambda$CDM cosmology, Bennett et al. \cite{18} obtained a substantially accurate result $H_0=69.6\pm0.7$ km s$^{-1}$ Mpc$^{-1}$. In succession, adopting other mid-redshift data including the BAO peak at $z=0.35$ \cite{19}, 18 $H(z)$ data points \cite{20,21,22}, 11 ages of old high-redshift galaxies \cite{23,24} and the angular diameter distance data from the Bonamente et al. galaxy cluster sample \cite{25}, Lima et al. got  $H_0=74.1\pm2.2$ km s$^{-1}$ Mpc$^{-1}$ \cite{26} in a $\Lambda$CDM model. Furthermore, replacing the Bonamente et al. galaxy cluster sample with the Filippis et al. one \cite{27}, Holanda et al. \cite{28} gave $H_0=70\pm4$ km s$^{-1}$ Mpc$^{-1}$. This implies that different mid-redshift data can provide different values of $H_0$. In addition, based on the fact that different observations should provide the same luminosity distance (LD) at a certain redsift, Wu et al. \cite{29} proposed a model-independent method to determine $H_0$. They obtained $H_0=74.1\pm2.2$ km s$^{-1}$ Mpc$^{-1}$ by combining the Union 2.1 SNe Ia data with galaxy cluster data \cite{30}.

Recently, the improved local measurement $H_0=73.24\pm1.74$ km s$^{-1}$ Mpc$^{-1}$ from Riess et al. 2016 \cite{31} exhibits a stronger tension with the Planck 2015 release $H_0=66.93\pm0.62$ km s$^{-1}$ Mpc$^{-1}$ \cite{32} at $3.4\sigma$ level. The improvements different from Riess et al. 2011 can be concluded as follows: (i) using new, near-infrared observations of Cepheid variables in 11 SNe Ia hosts; (ii) increasing the sample size of ideal SNe Ia calibrators from 8 to 19; (iii)  giving the calibration for a magnitude¨Credshift relation based on 300 SNe Ia at $z<0.15$; (iv) a $33\%$ reduction of the systematic uncertainty in the maser distance to the NGC 4258; (v) a more robust distance to the Large Magellanic Cloud (LMC) based on the late-type detached eclipsing binaries (DEBs); (vi) increasing the sample size of Cepheids in the LMC; (vii) Hubble Space Telescope (HST) observations of Cepheids in M31; (viii) using new HST-based trigonometric parallaxes for the MW Cepheids.

Due to the stronger tension than before between the local and global measurement of $H_0$, we would like to derive the value of $H_0$ using the latest 36 $H(z)$ data points (see Table. \ref{t1}) based on the model-independent method---Gaussian Processes (GP). In 2014, Busti et al. \cite{33} utilized the GP method to derive the the value of $H_0$ and obtained $H_0=64.9\pm4.2$ km s$^{-1}$ Mpc$^{-1}$, which is very consistent with the Planck 2015 analysis but still exists a $1.8\sigma$ tension with the Riess et al. 2016 result. It is worth noting that they use only 19 H(z) data points of passively evolving galaxies as cosmic chronometers \cite{34}.
To date, we expect to use the more and higher-quality $H(z)$ data than before to derive the value of $H_0$ by using the nonparametric GP method.

This study is organized in the following manner. In Section 2, we perform our methodology briefly. In Section 3, the GP reconstruction results are exhibited. In Section 4, we simulate the same quality $H(z)$ data as today to forecast the future constraints on $H_0$. The discussions and conclusions are presented in the final section.

\section{The Methodology}
Generally speaking, the GP exhibits a distribution over functions, and is a generalization of a Gaussian distribution which is the distribution of a random variable. The GP algorithm is a fully Bayesian approach for smoothing data, and can be used to implement a reconstruction of a function directly from data without assuming a concrete parameterization of the function. Consequently, one can determine any cosmological quantity directly from the correspondingly cosmic data, and the key requirement of the GP algorithm is only the covariance function which entirely depends on the observed cosmological data. At each point $x$, the reconstructed function $f(x)$ is a Gaussian distribution with a mean value and Gaussian error. The key of the GP is a covariance function $k(x,\tilde{x})$ which correlates the function $f(x)$ at different reconstruction points. The covariance function $k(x,\tilde{x})$ depends only on two hyperparameters $l$ and $\sigma_f$, which describe the coherent scale of the correlation in $x$-direction and typical change in the $y$-direction, respectively. Due to this special advantage, the GP has been widely applied for different purposes in the literature: investigating the expansion dynamics of the universe \cite{45,46,47}, the distance duality relation \cite{48}, the cosmography \cite{49}, the test of the $\Lambda$CDM model \cite{50}, the determination of the interaction between dark energy and dark matter \cite{51}, etc.

In the present analysis, we use the the public package GaPP (Gaussian Processes in Python) \cite{52} to carry out the reconstruction, which is firstly invented by Seikel et al. In the meanwhile, we take into account the squared exponential covariance function (SECF)
\begin{equation}
k(x,\tilde{x})=\sigma_f^2 exp[-\frac{(x-\tilde{x})^2}{2l^2}]. \label{1}
\end{equation}
Furthermore, we also consider three parametric models in order to compare the results obtained by the GP method with standard analysis. At first, we consider a flat $\omega$CDM model and the corresponding Hubble parameter is expressed as
\begin{equation}
H_{\omega}(z)= H_0\{\Omega_{m0}(1+z)^3+(1-\Omega_{m0})(1+z)^{3(1+\omega)}\}^{\frac{1}{2}}, \label{2}
\end{equation}
where $\omega$ and $\Omega_{m0}$ denote the dark energy equation of state and the matter density ratio parameter, respectively. It is clearly that the $\omega$CDM model will reduce to the flat $\Lambda$CDM model when $\omega=-1$. Subsequently, we also consider the popular decaying vacuum model \cite{53} whose Hubble parameter can be written as
\begin{equation}
H_{D}(z)= H_0\{\frac{3\Omega_{m0}}{3-\epsilon}(1+z)^{3-\epsilon}+1-\frac{3\Omega_{m0}}{3-\epsilon}\}^{\frac{1}{2}} \label{2},
\end{equation}
where $\epsilon$ is a small positive constant characterizing the deviation from the standard matter expansion rate.

\section{The results}

\begin{figure}
\centering
\includegraphics[scale=0.3]{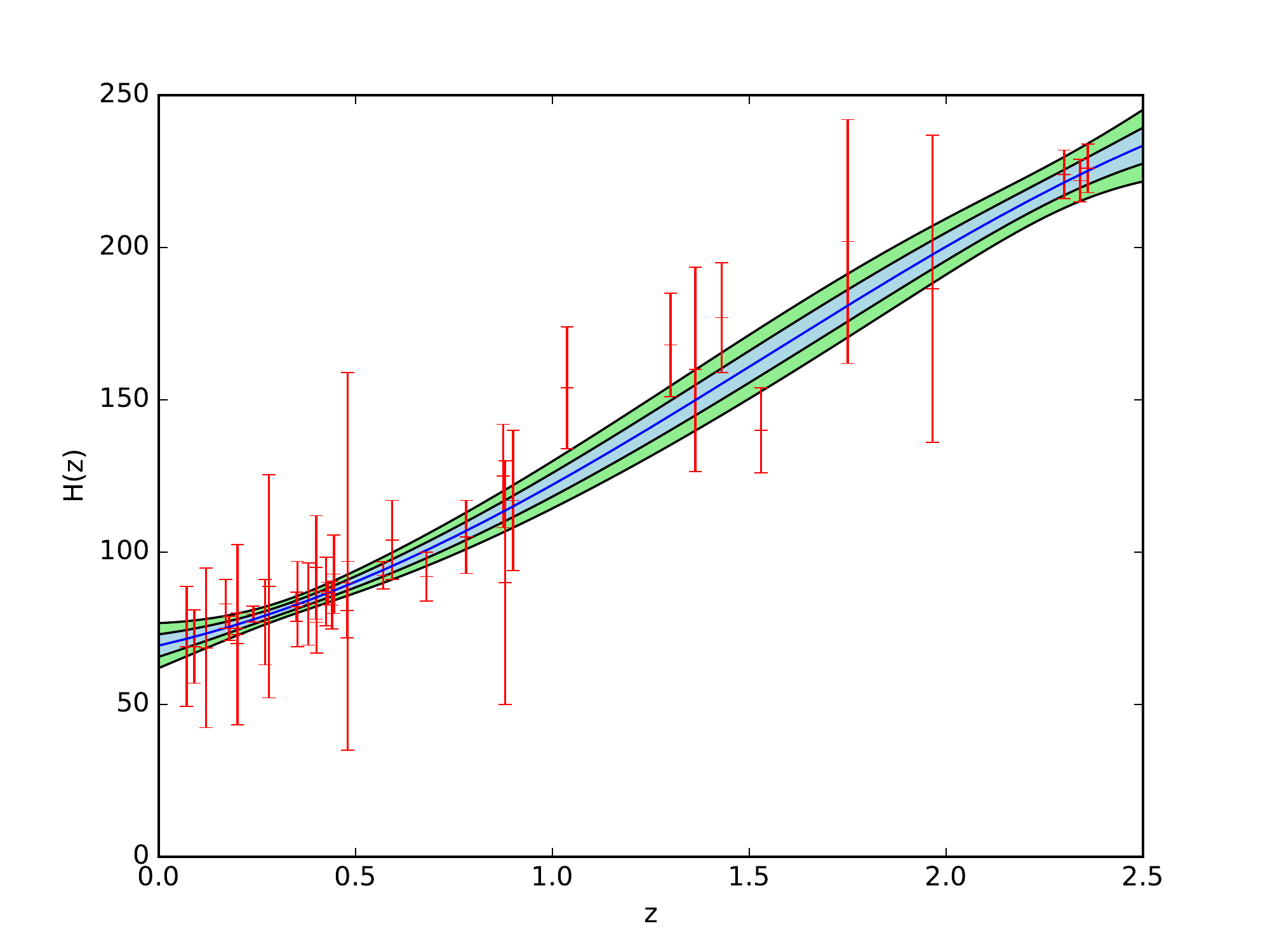}
\includegraphics[scale=0.3]{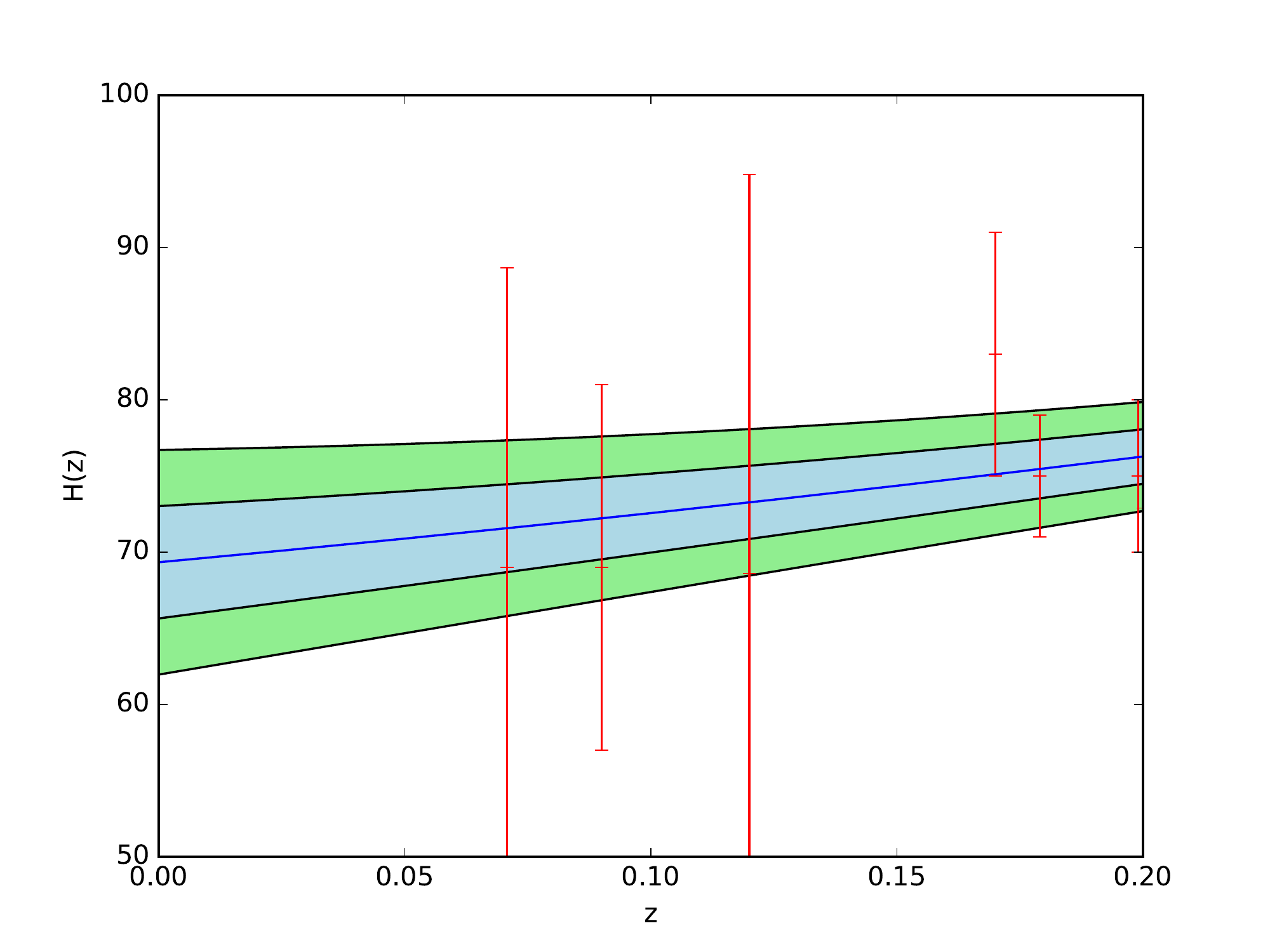}
\caption{In the left panel, we exhibit the GP reconstruction of $H(z)$ using the 36 $H(z)$ data points. To show our reconstruction result on $H_0$ better, we also plot for a small redshift range in the right panel. The blue lines represent the mean value of the reconstruction. The shaded regions are reconstructions with $68\%$ and $95\%$ confidence level.}\label{f1}
\end{figure}
\begin{figure}
\centering
\includegraphics[scale=0.5]{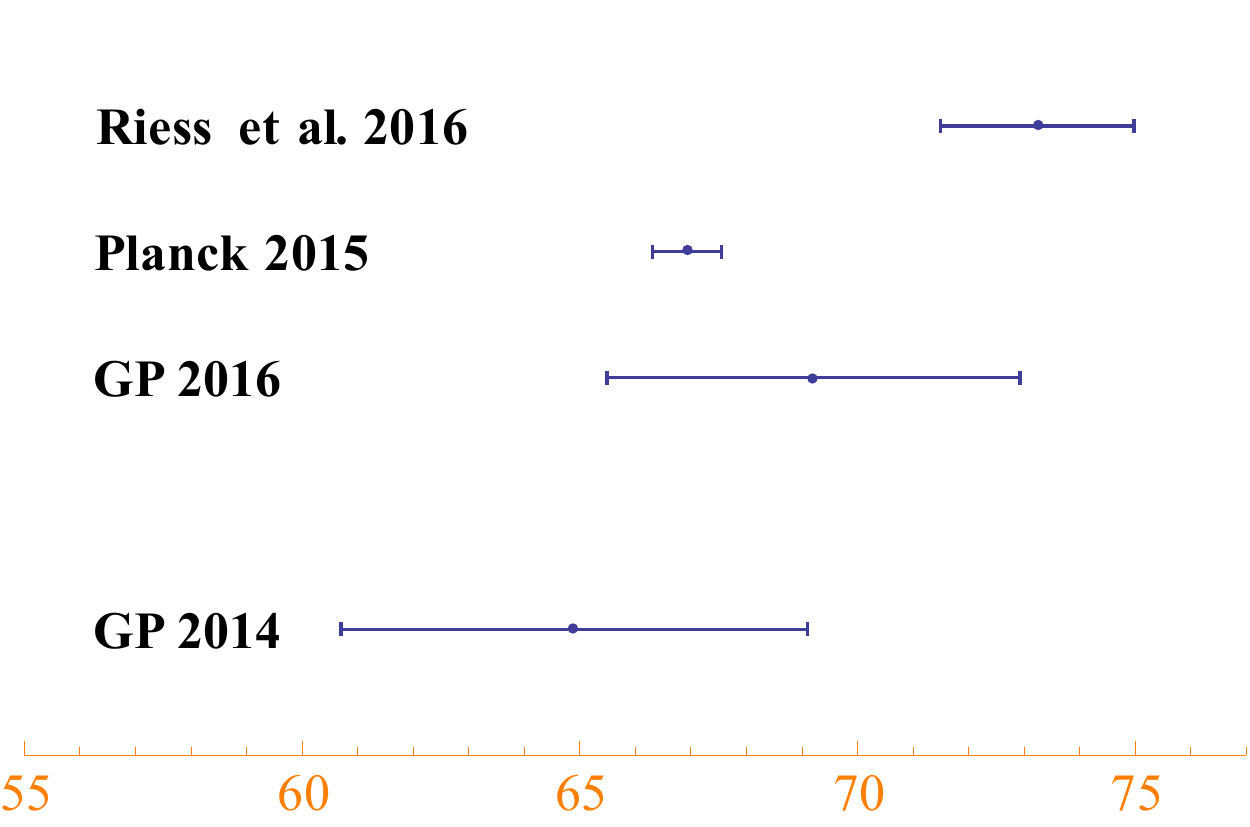}
\caption{Constraints on the Hubble constant $H_0$.}\label{f2}
\end{figure}
\begin{figure}
\centering
\includegraphics[scale=0.3]{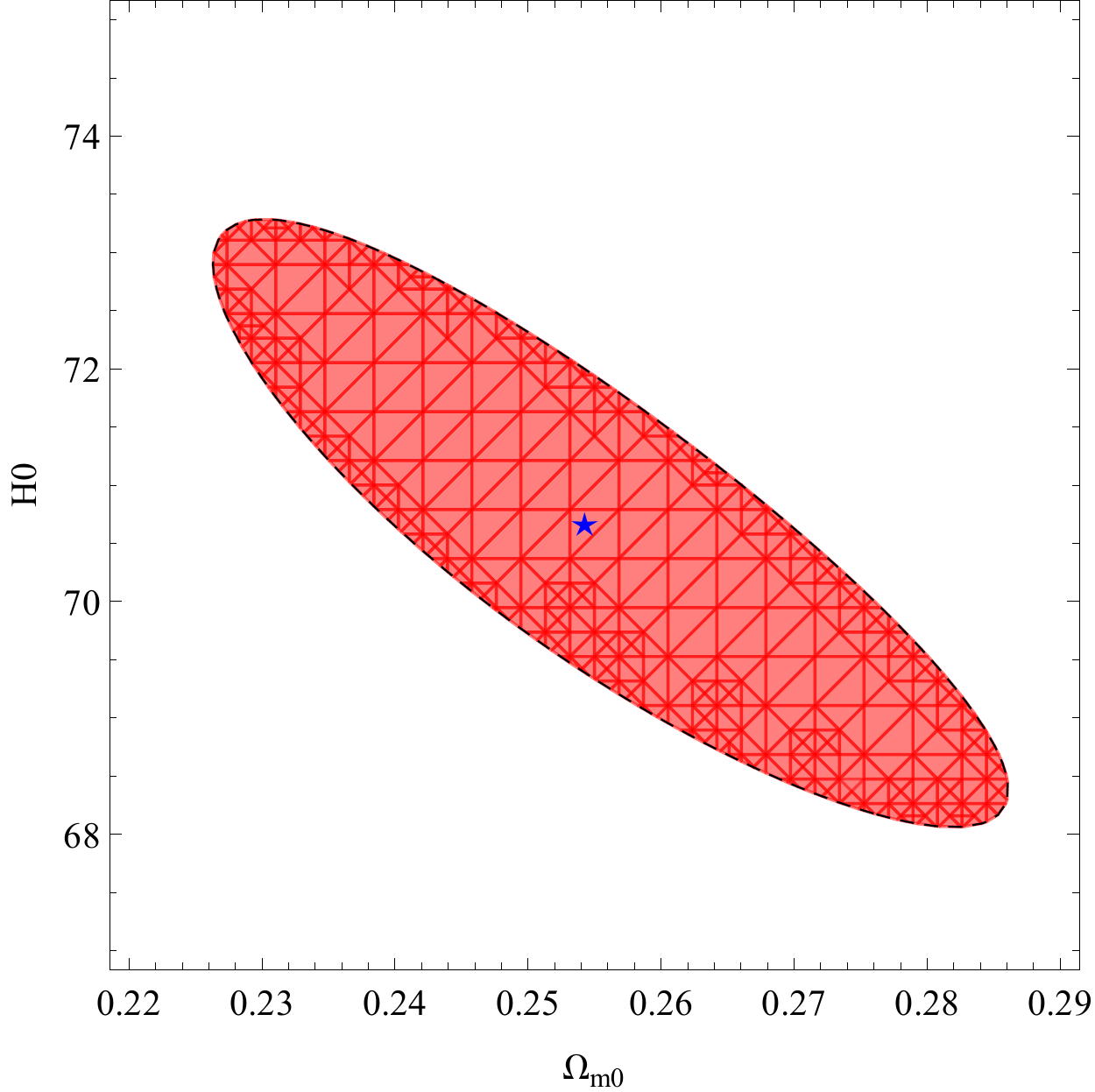}
\includegraphics[scale=0.3]{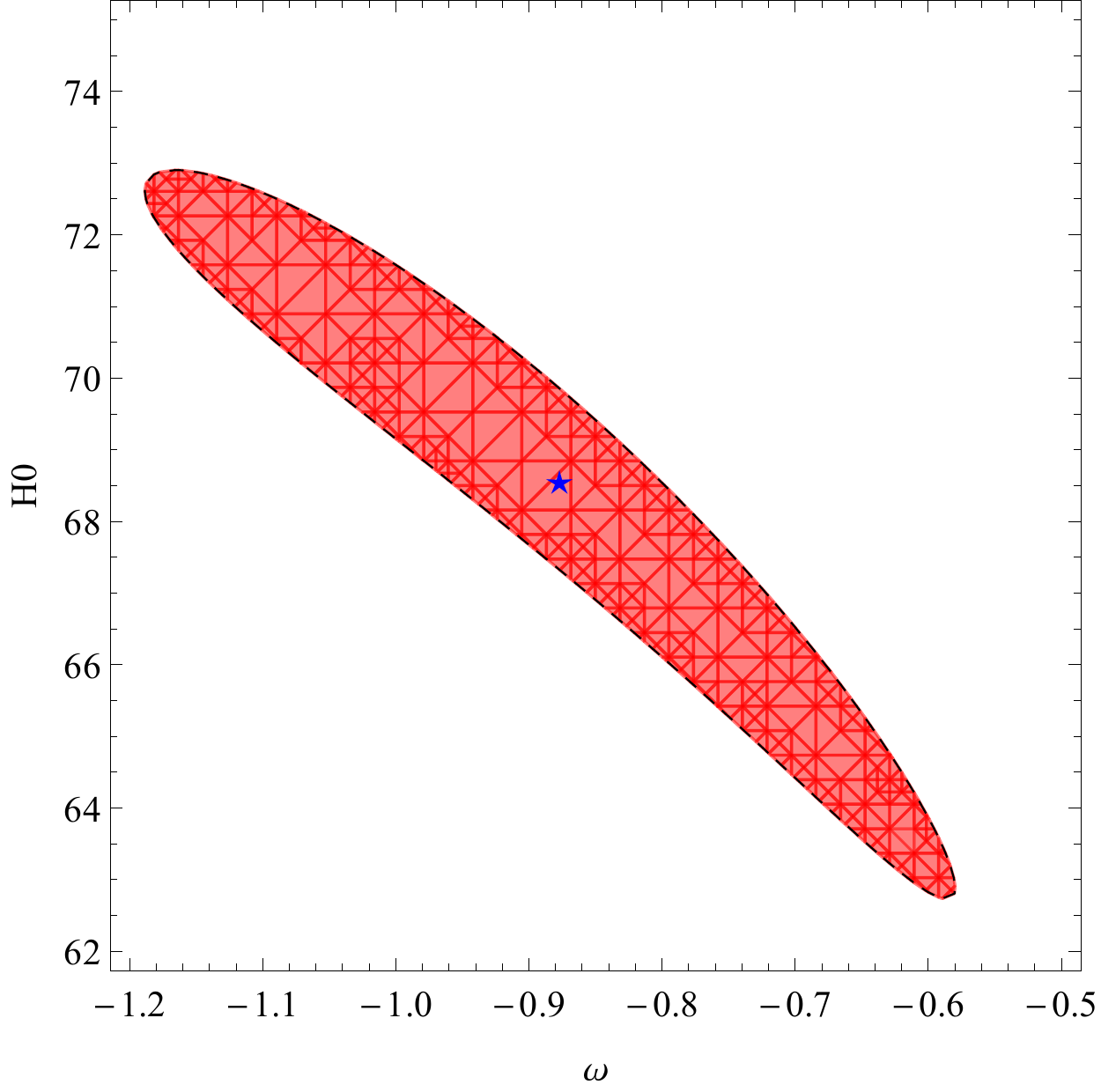}
\includegraphics[scale=0.3]{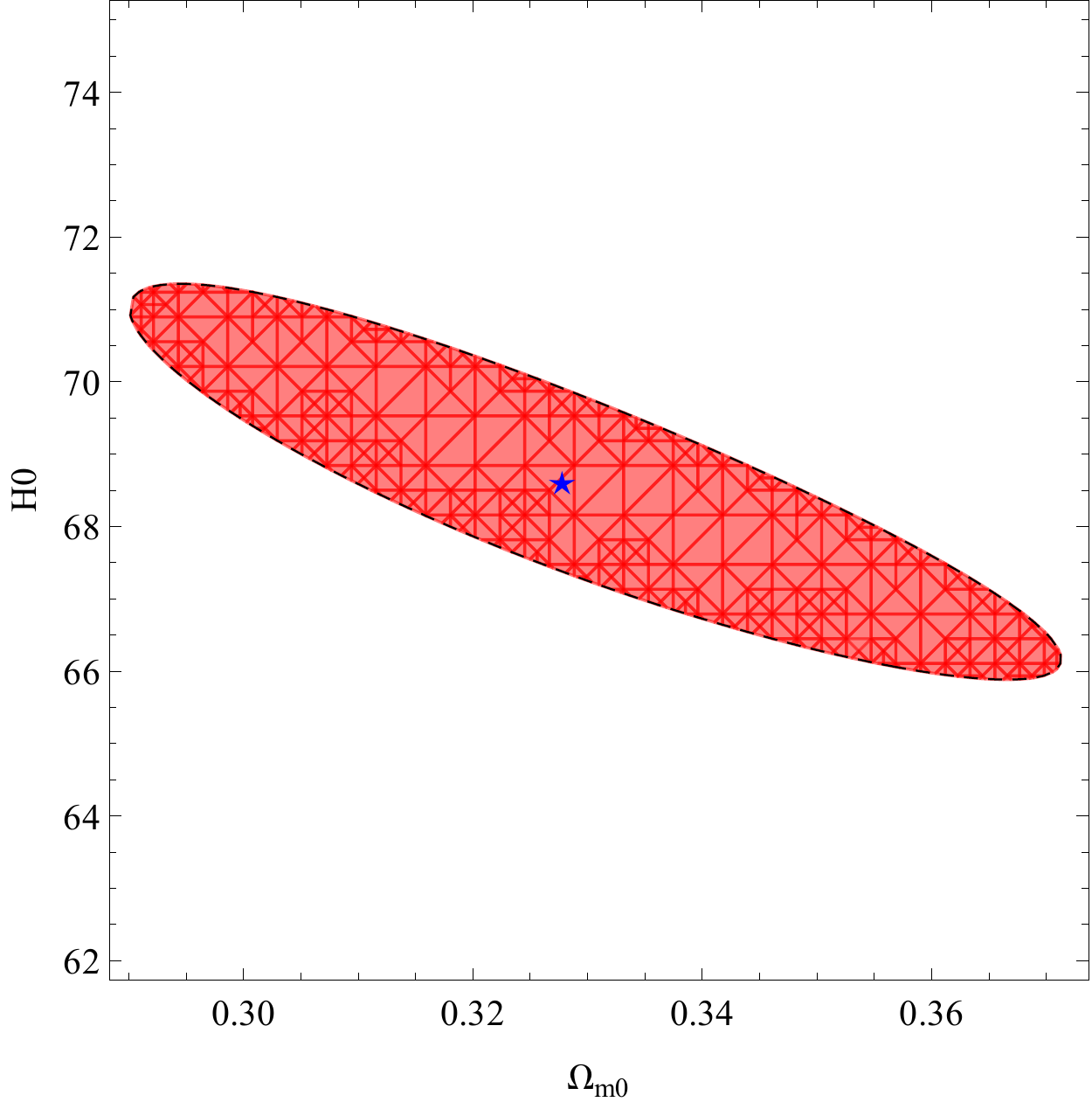}
\caption{The two-dimensional likelihoods from left to right for the $\Lambda$CDM model parameter pair $(\Omega_{m0},H_0)$, $\omega$CDM model parameter pair $(\omega,H_0)$ and decaying vacuum model parameter pair $(\Omega_{m0},H_0)$, respectively. The labels $`` \star "$ denote the best fitting points.}\label{f3}
\end{figure}
\begin{figure}
\centering
\includegraphics[scale=0.3]{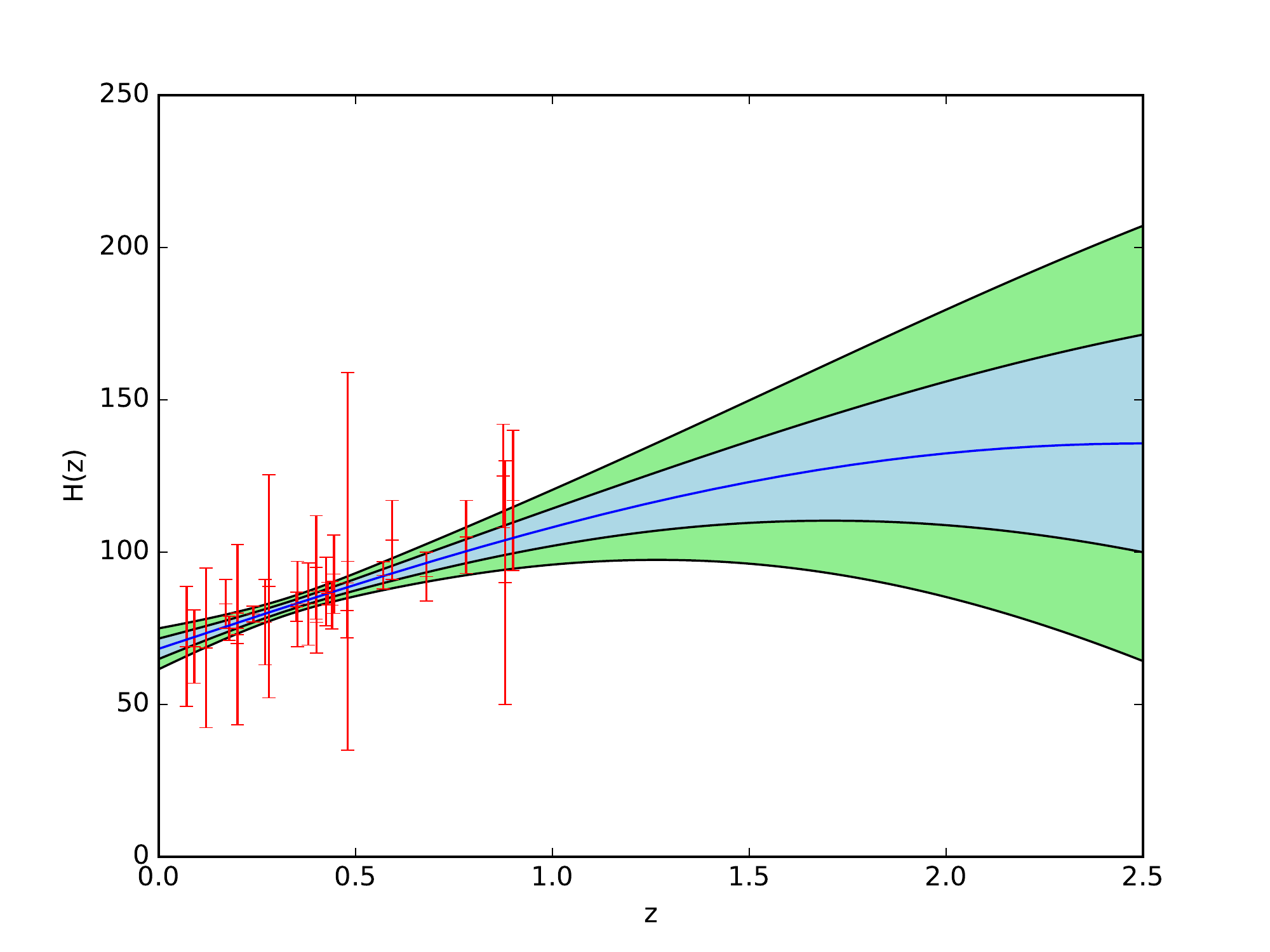}
\includegraphics[scale=0.3]{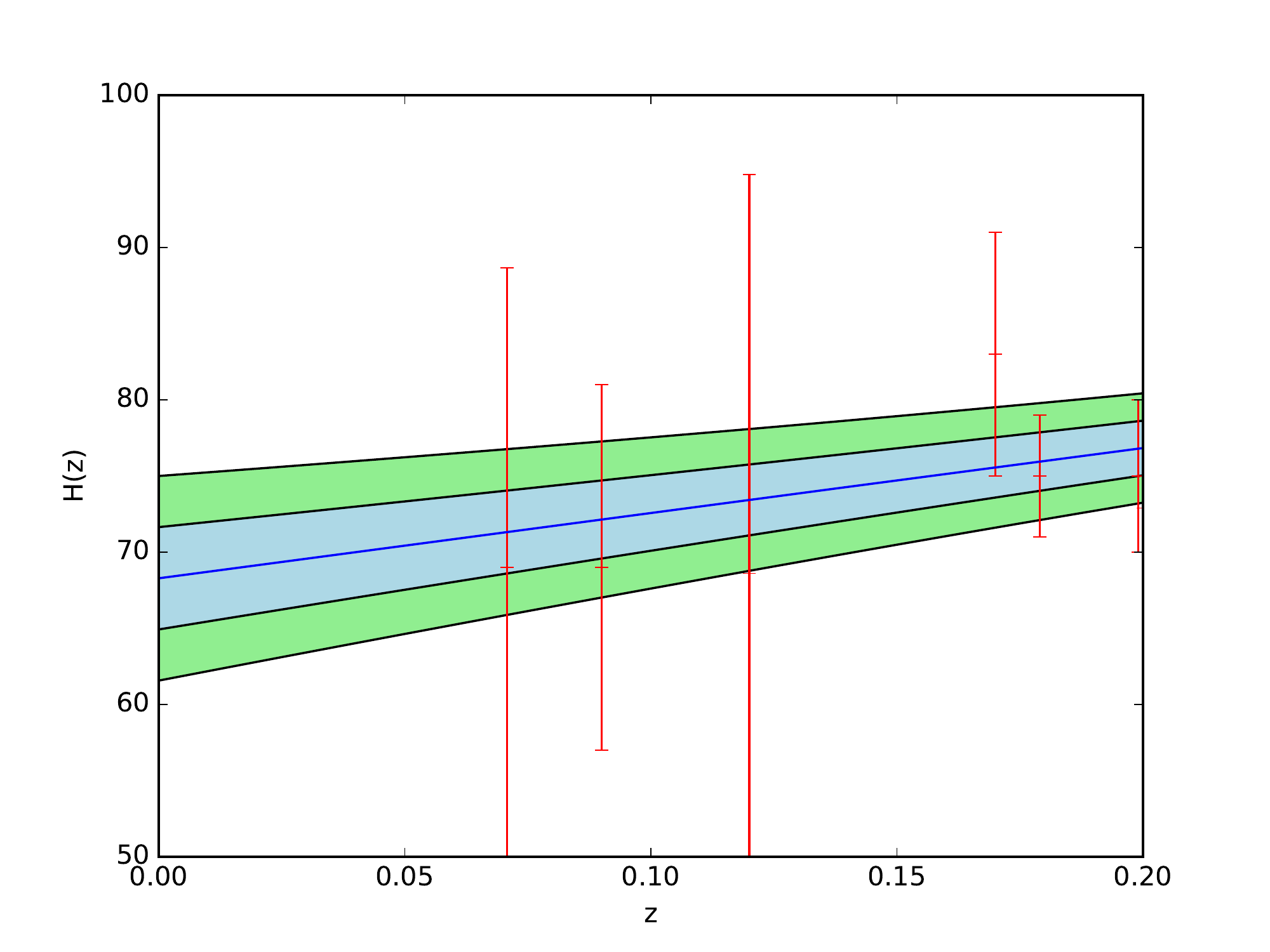}
\includegraphics[scale=0.3]{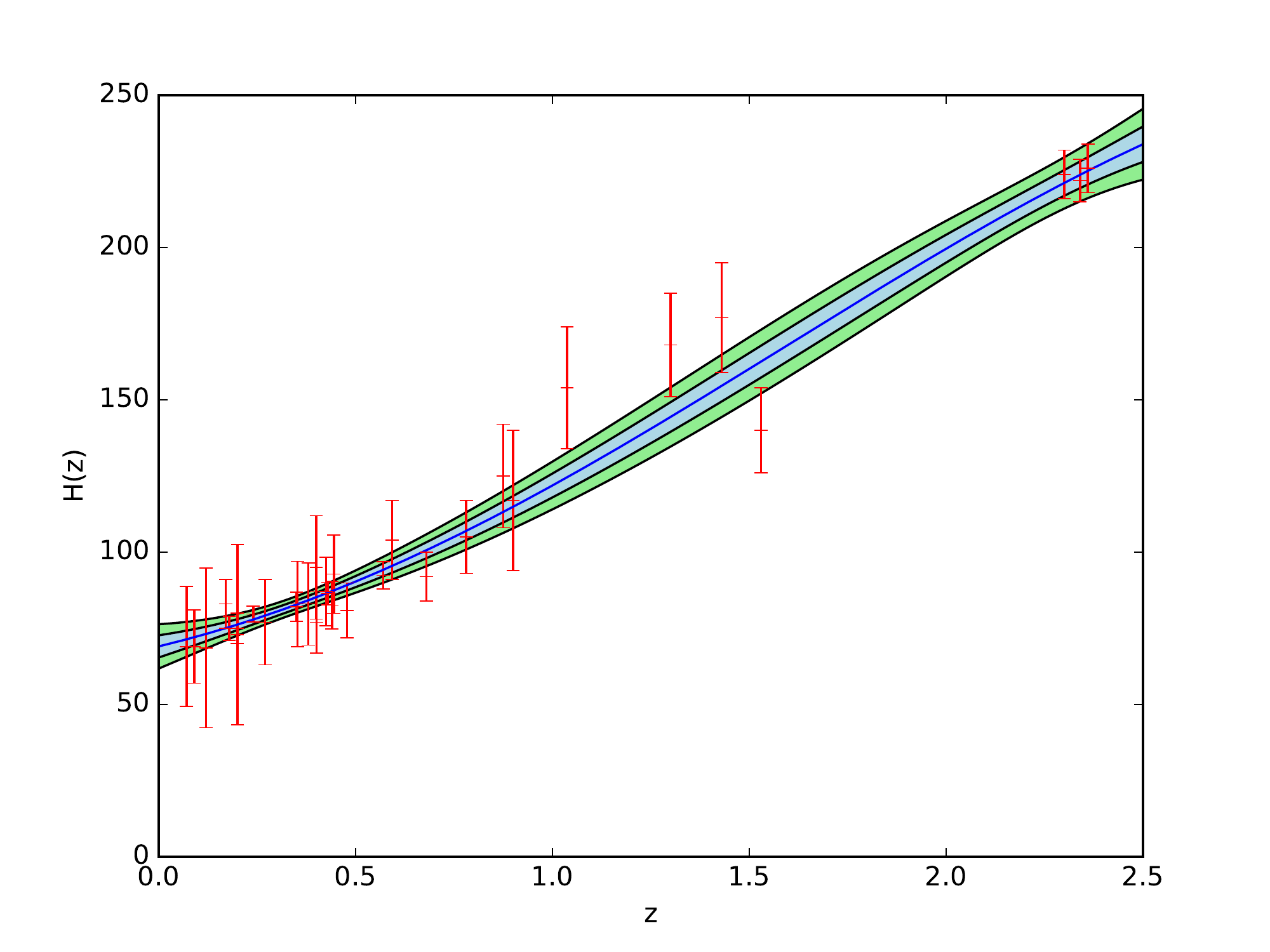}
\includegraphics[scale=0.3]{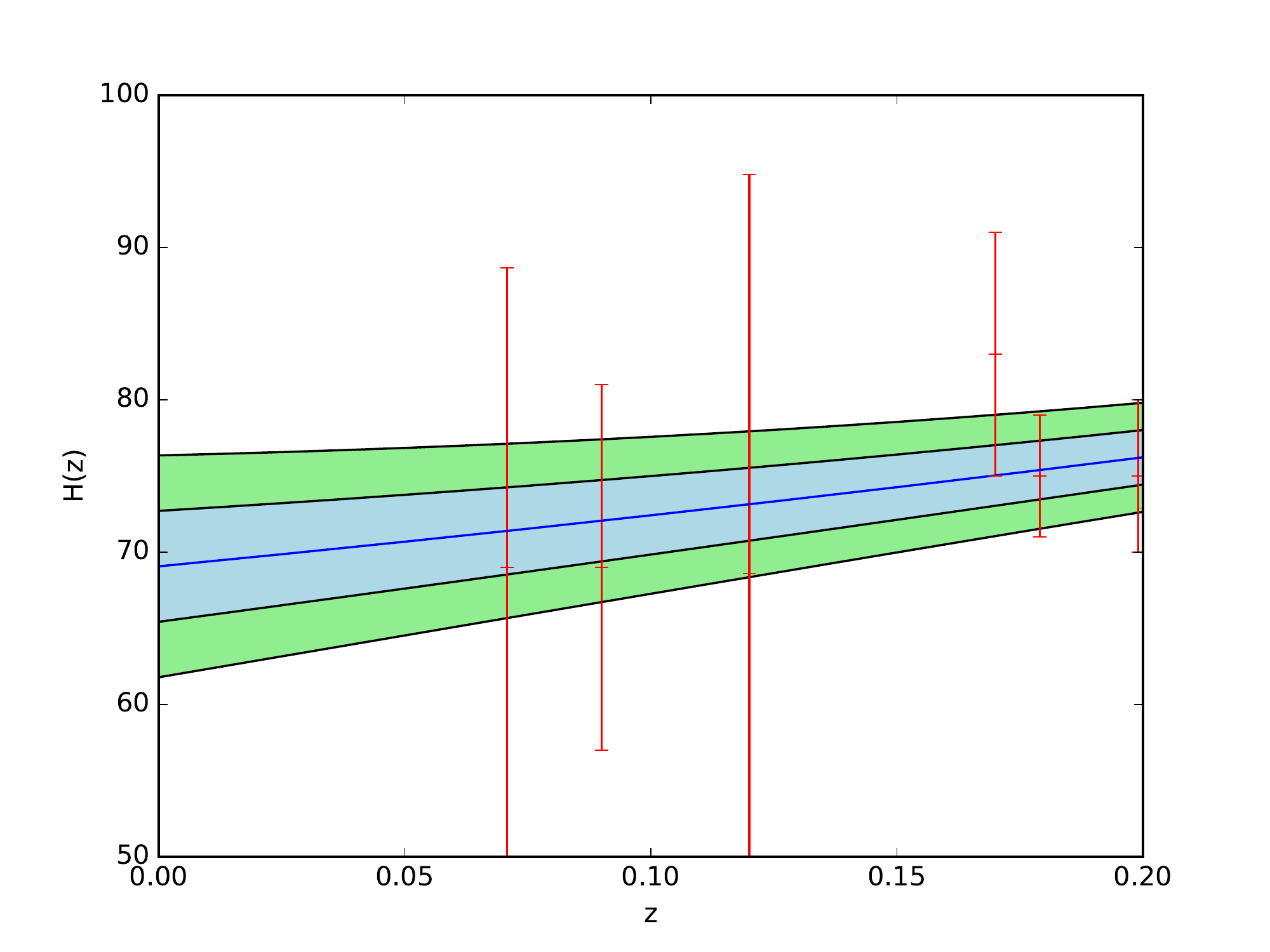}
\caption{In the upper panels, we exhibit the GP reconstruction of $H(z)$ removing the data points with redshifts $z>1$ from 36 $H(z)$ data points. In the lower panels, we exhibit the GP reconstruction of $H(z)$ removing the data points with errors greater than 30 km s$^{-1}$ Mpc$^{-1}$.}\label{f4}
\end{figure}
\begin{figure}
\centering
\includegraphics[scale=0.3]{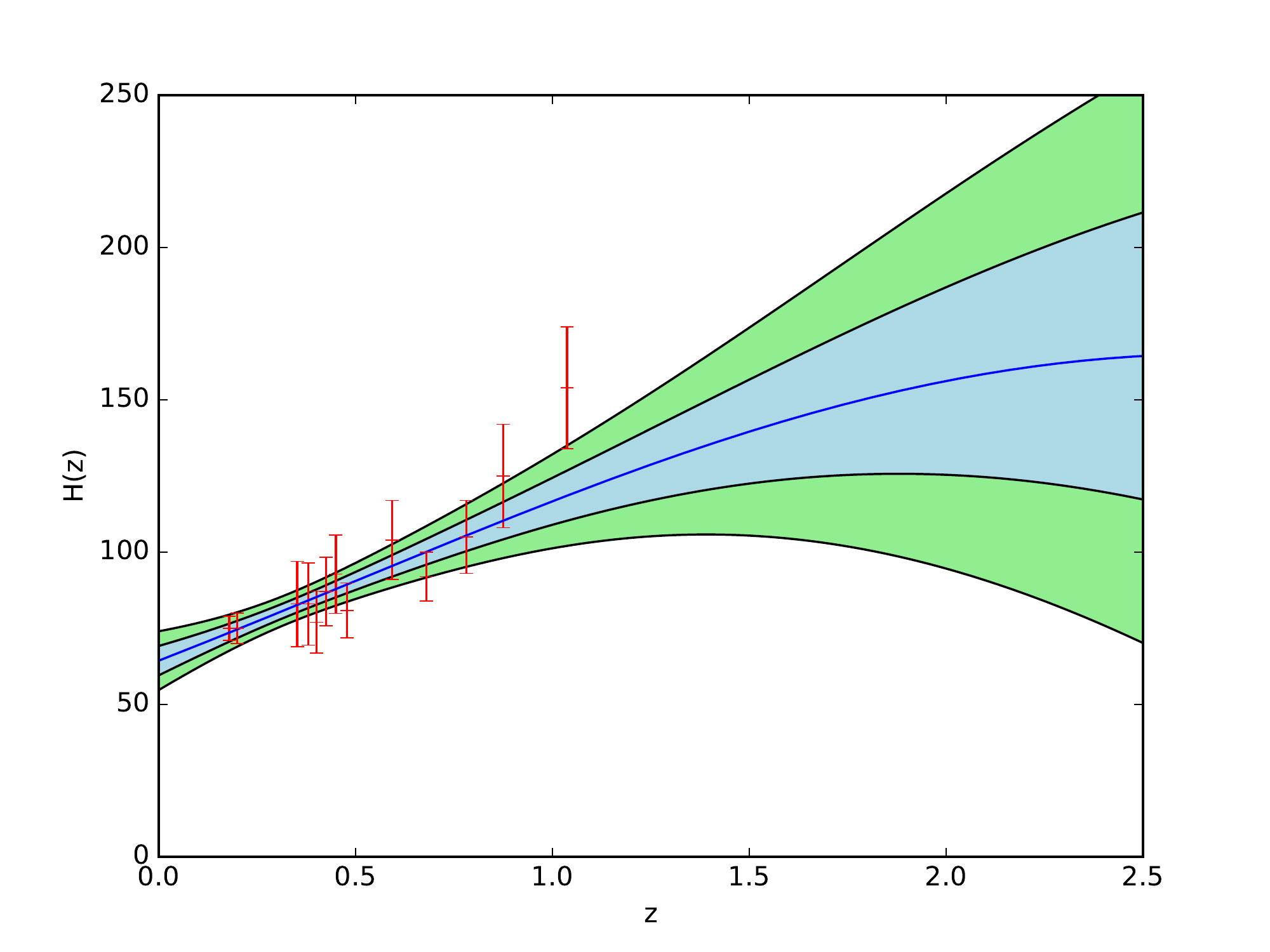}
\includegraphics[scale=0.3]{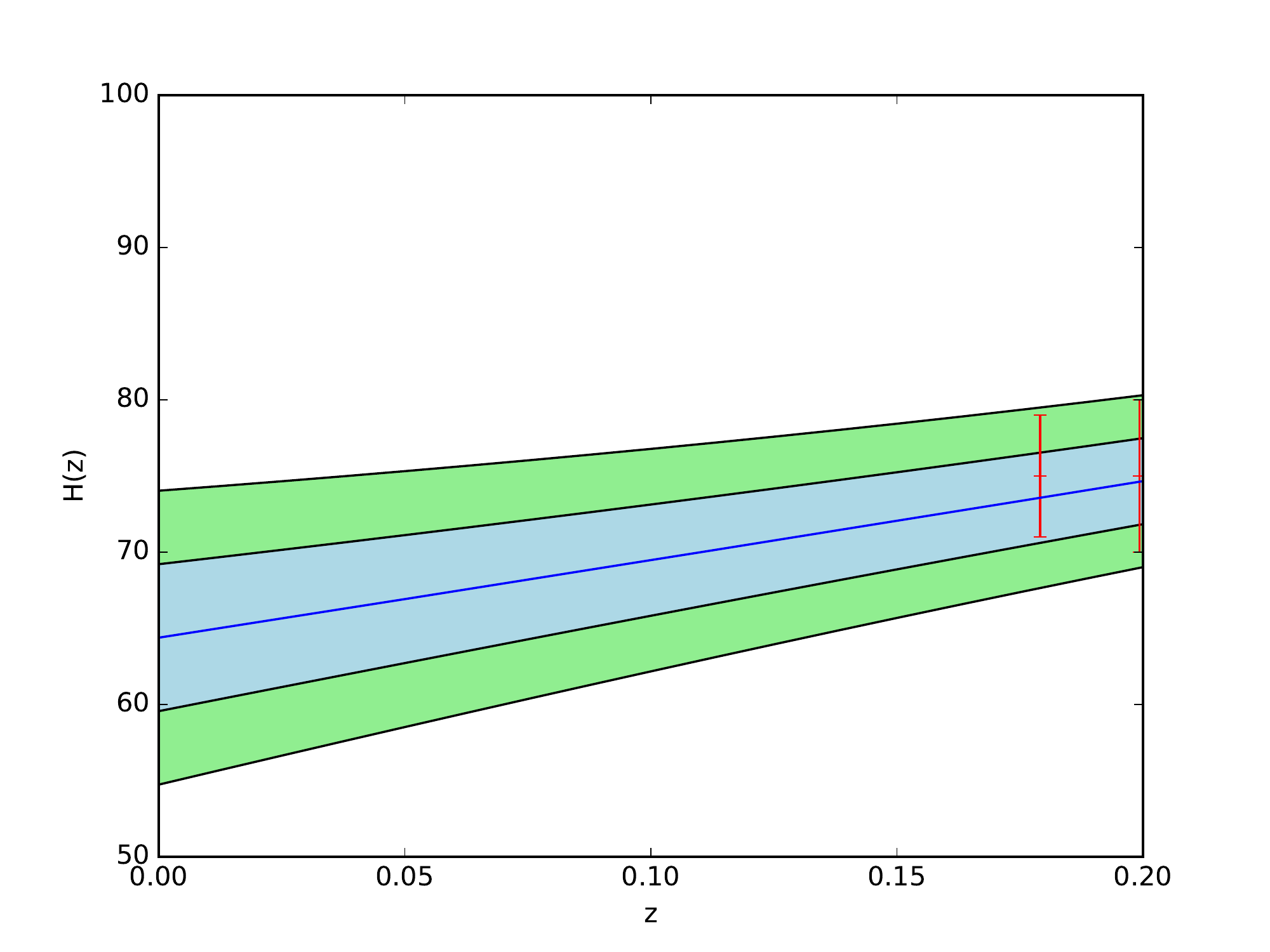}
\includegraphics[scale=0.3]{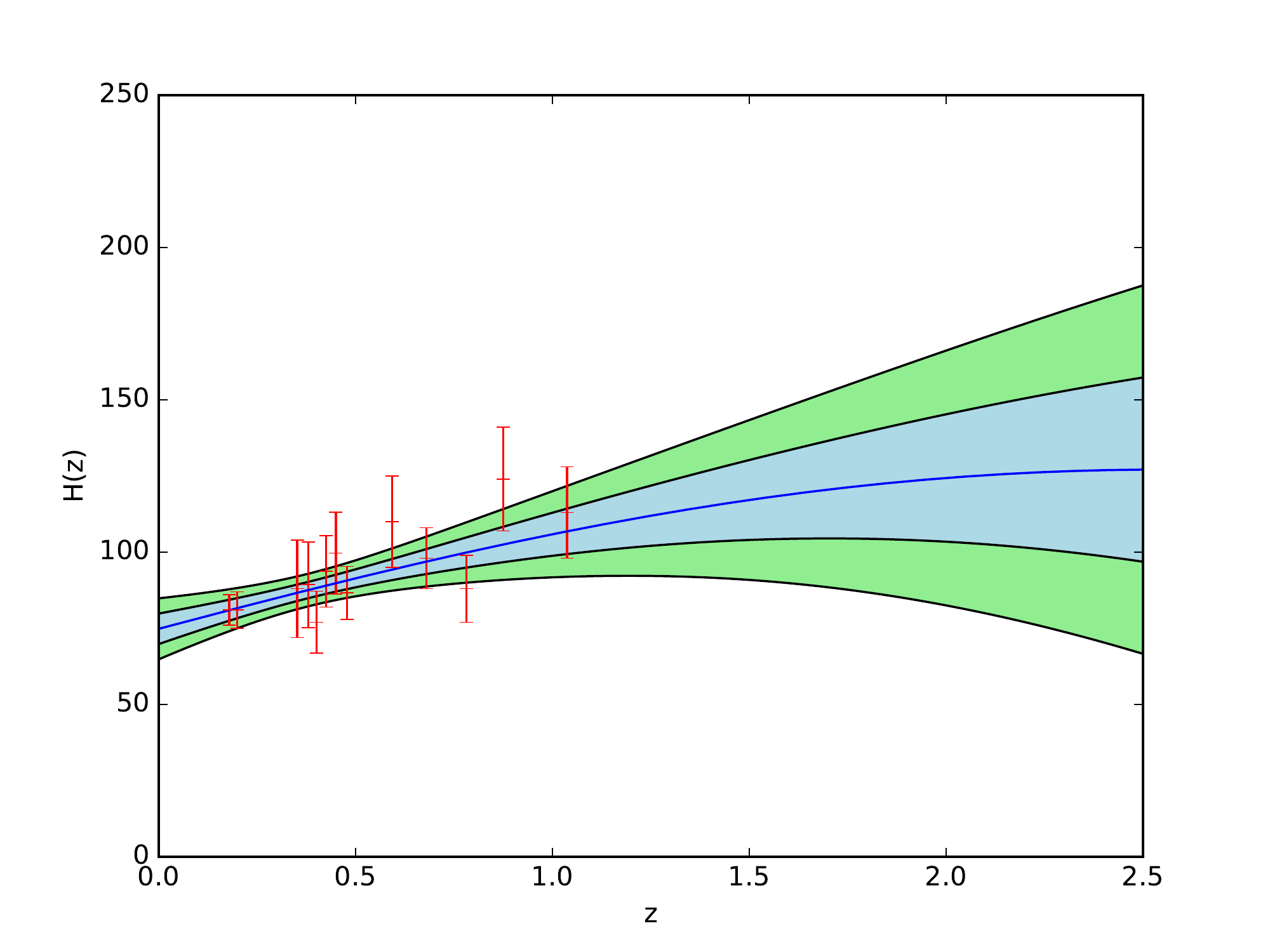}
\includegraphics[scale=0.3]{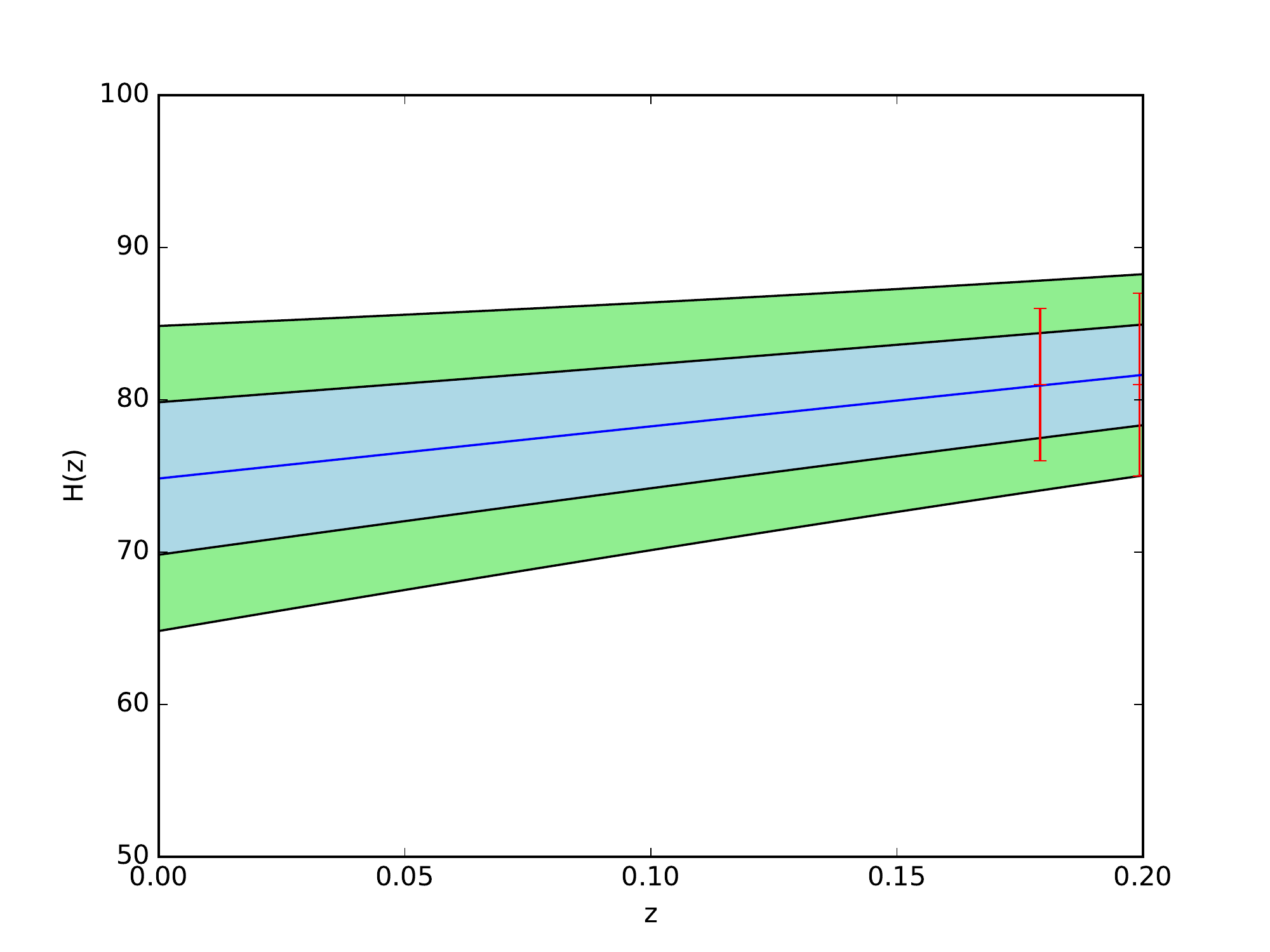}
\caption{From top to bottom, we exhibit the GP reconstructions of $H(z)$ using the 13 measurements from Moresco et al. [12,16] for BC03 and M11 models, respectively.}\label{f5}
\end{figure}
\begin{figure}
\centering
\includegraphics[scale=0.3]{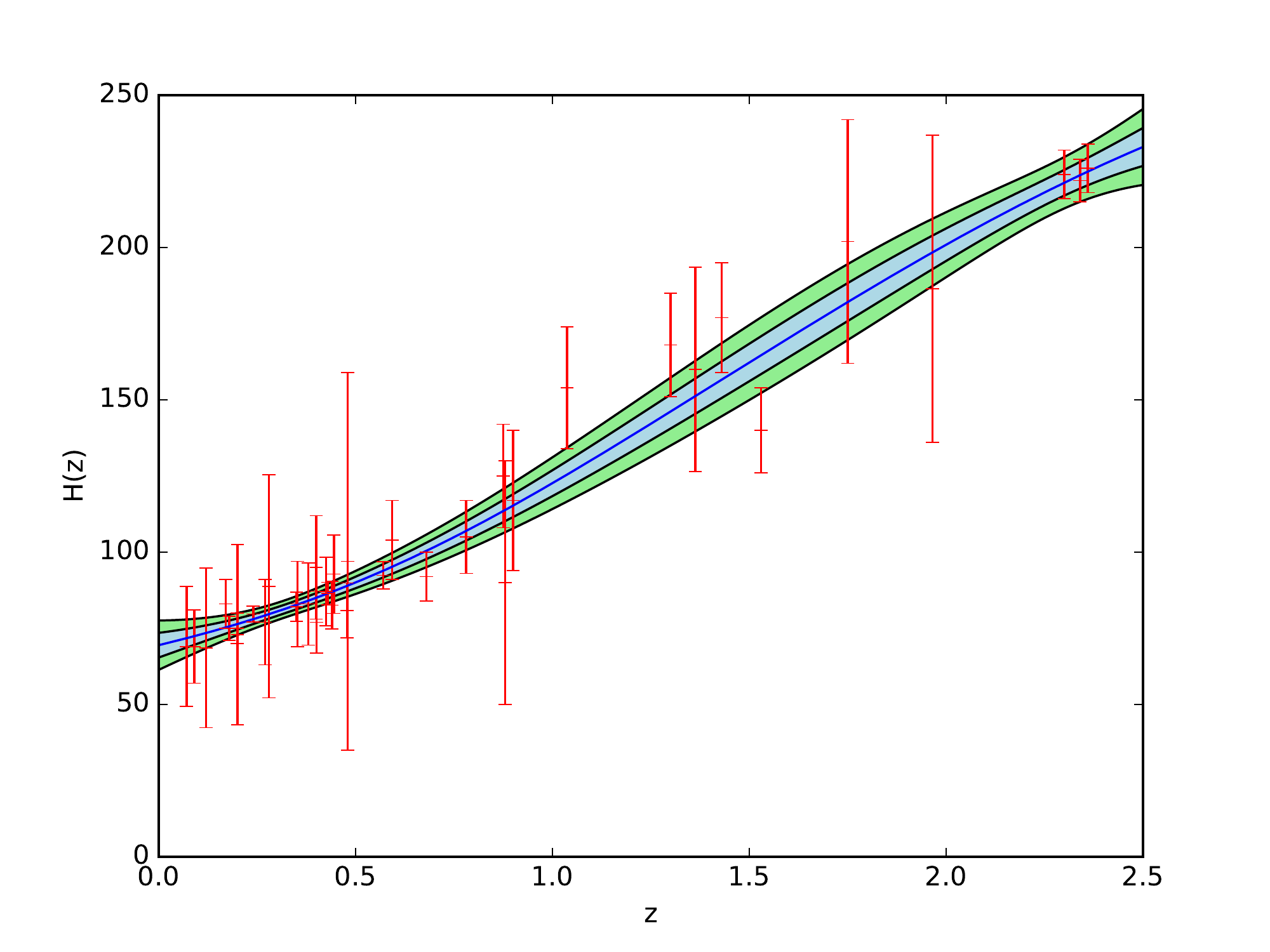}
\includegraphics[scale=0.3]{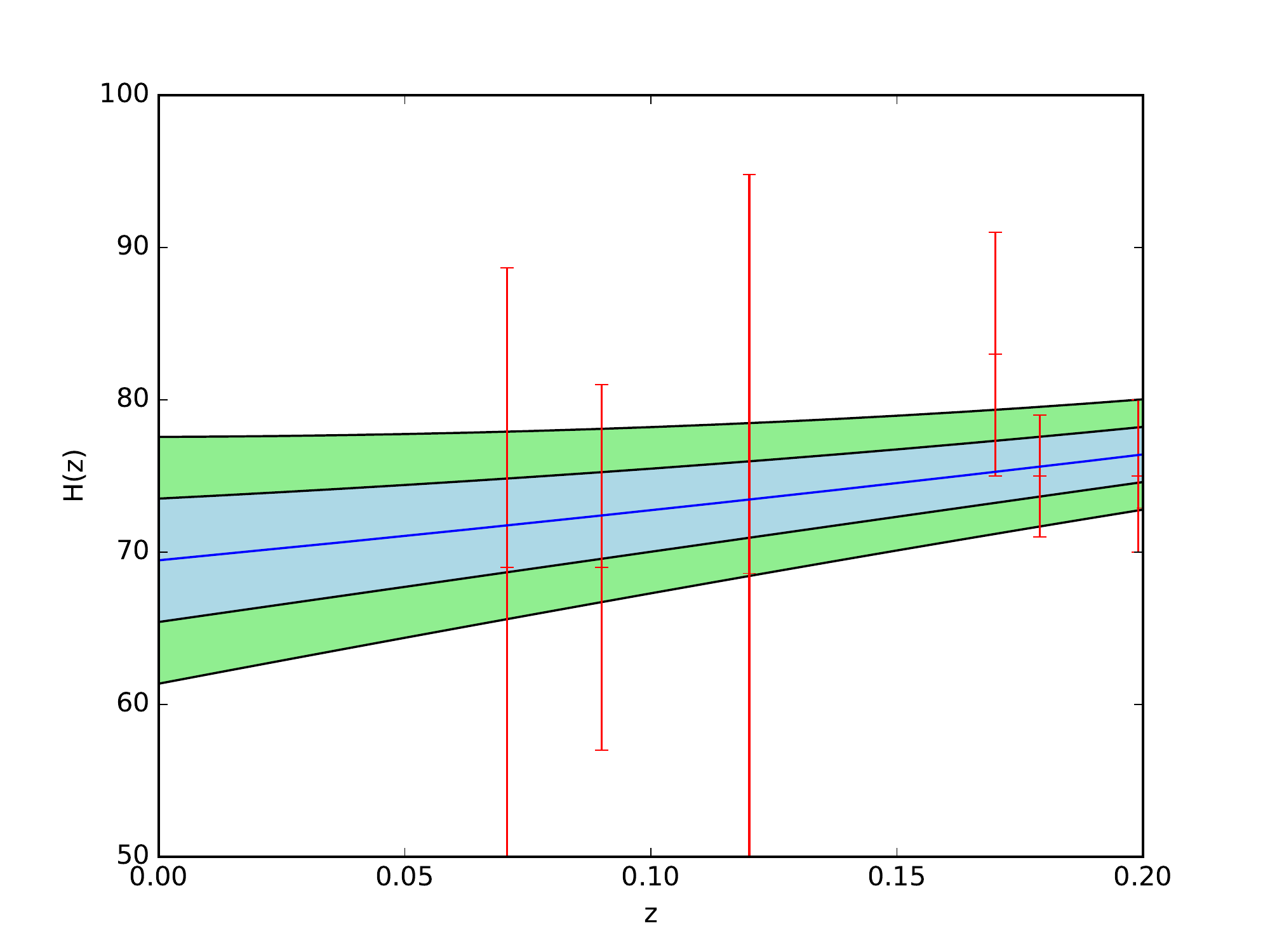}
\includegraphics[scale=0.3]{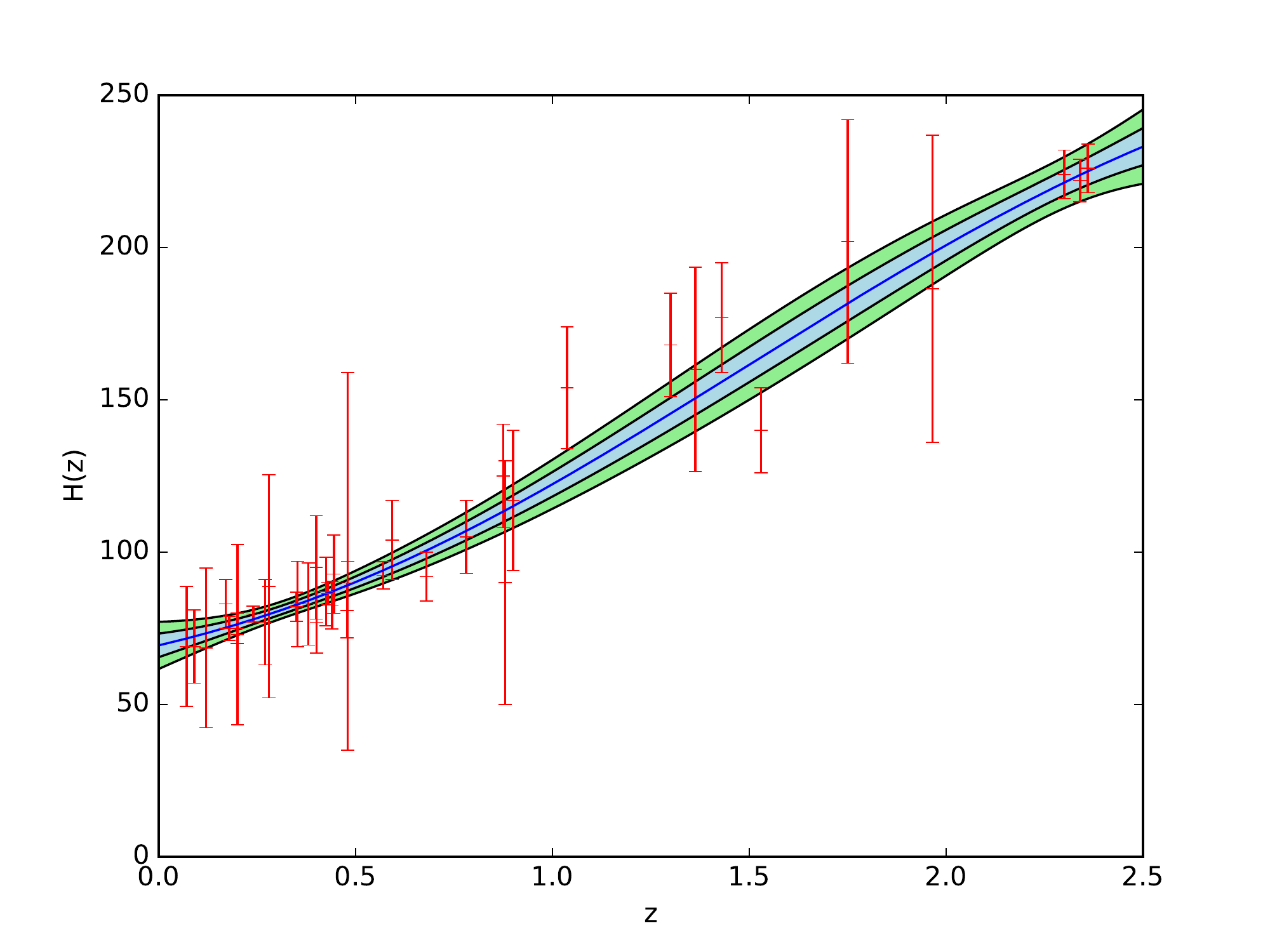}
\includegraphics[scale=0.3]{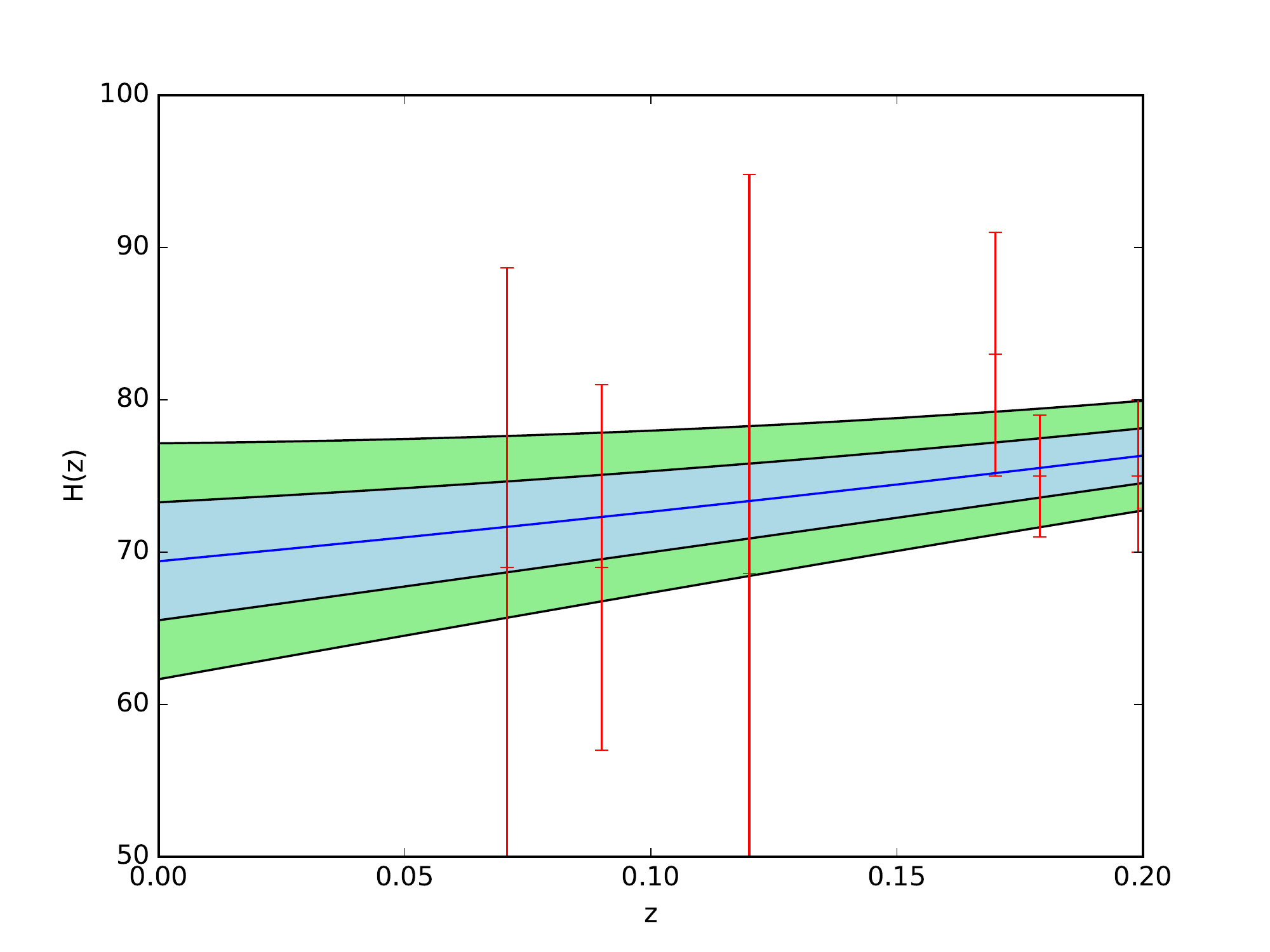}
\includegraphics[scale=0.3]{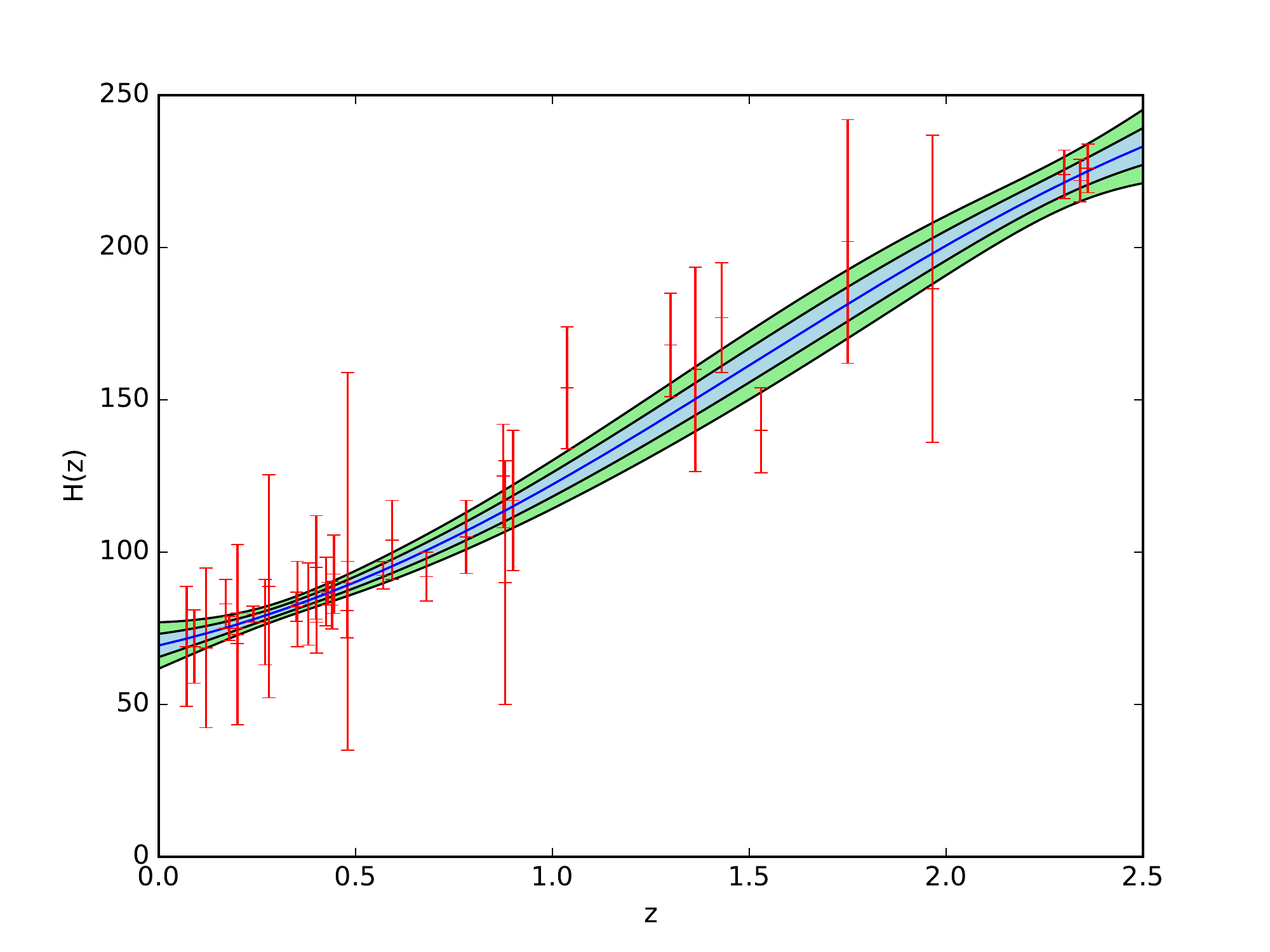}
\includegraphics[scale=0.3]{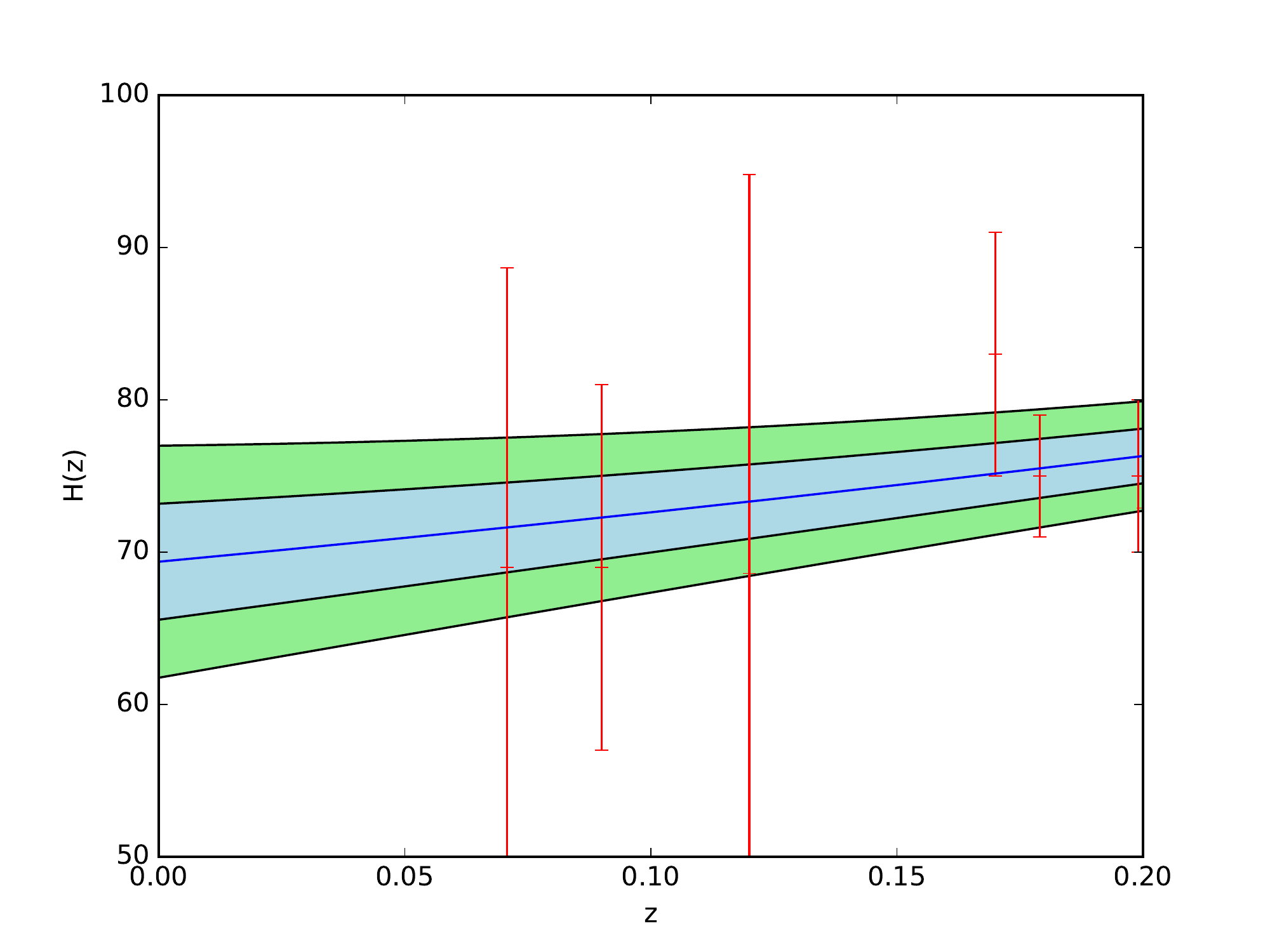}
\includegraphics[scale=0.3]{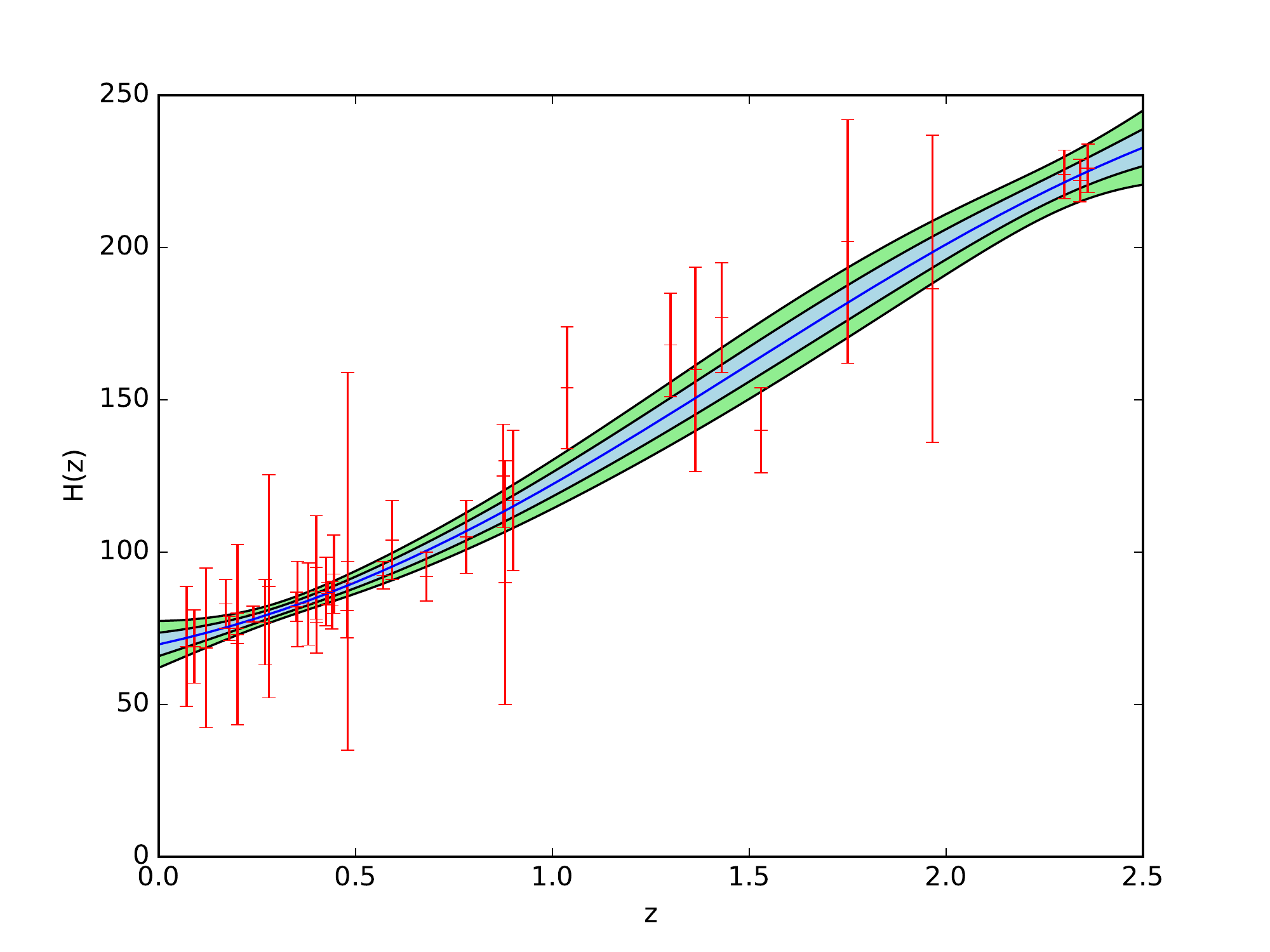}
\includegraphics[scale=0.3]{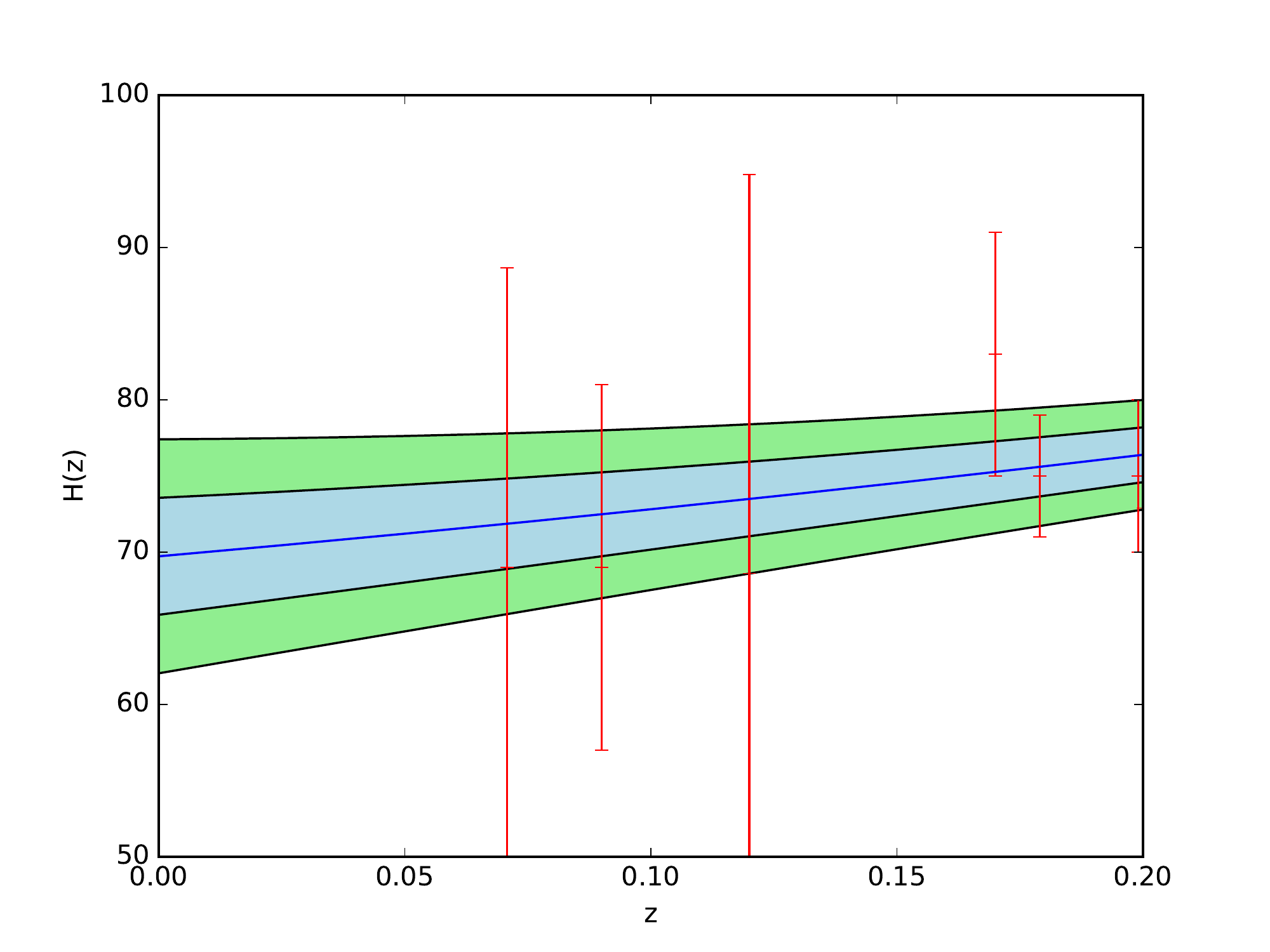}
\caption{From top to bottom, we exhibit the GP reconstructions of $H(z)$ using the 36 $H(z)$ data points for the covariance functions Mat\'{e}rn $(5/2)$, Mat\'{e}rn $(7/2)$, Mat\'{e}rn $(9/2)$ and Cauchy, respectively.}\label{f6}
\end{figure}
\begin{figure}
\centering
\includegraphics[scale=0.5]{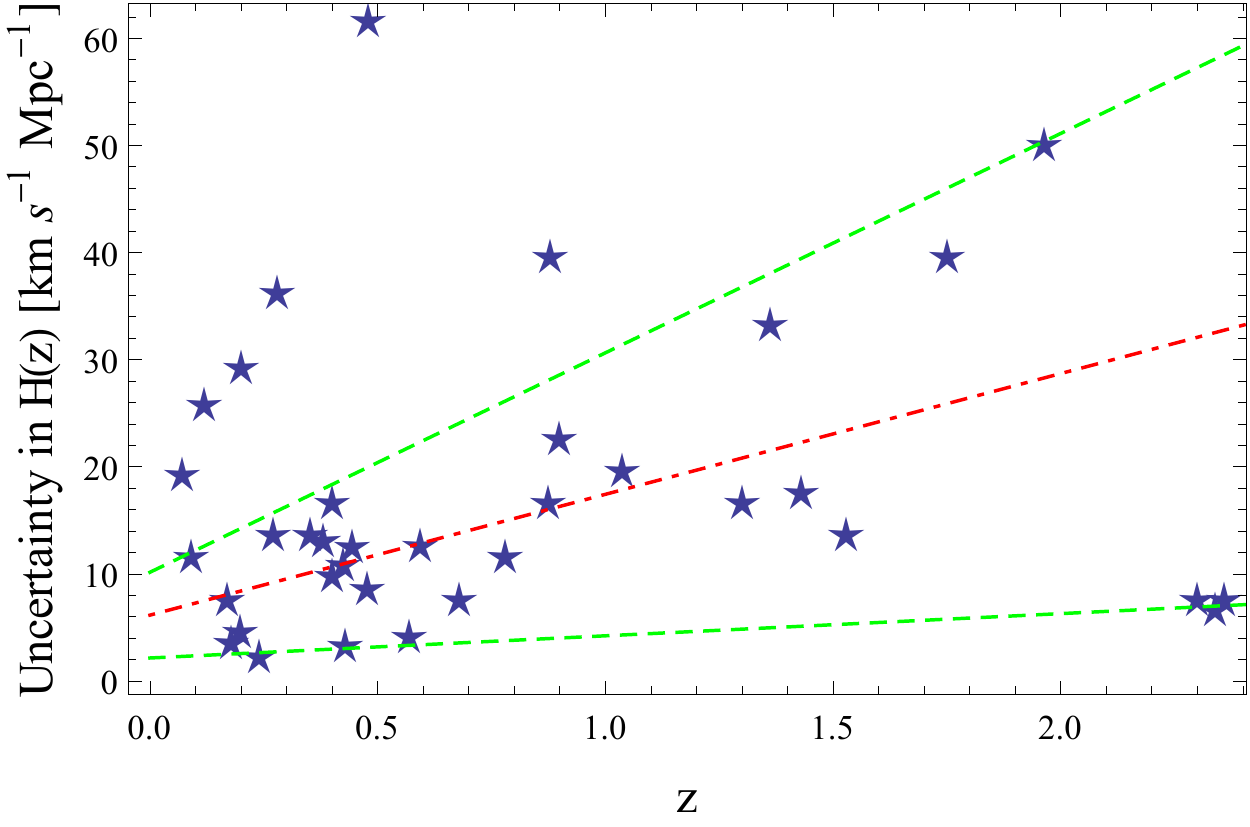}
\caption{The uncertainties of the extended 36 $H(z)$ measurements. The labels $\blacktriangle$ and $\star$ correspond to outliers and non-outliers, respectively. The two bounds $\sigma_{+}(z)$ and $\sigma_{-}(z)$ are plotted as two green (dashed) lines. The red (dash-dotted) line corresponds to the mean uncertainty $\sigma_{0}$.}\label{f7}
\end{figure}
\begin{figure}
\centering
\includegraphics[scale=0.3]{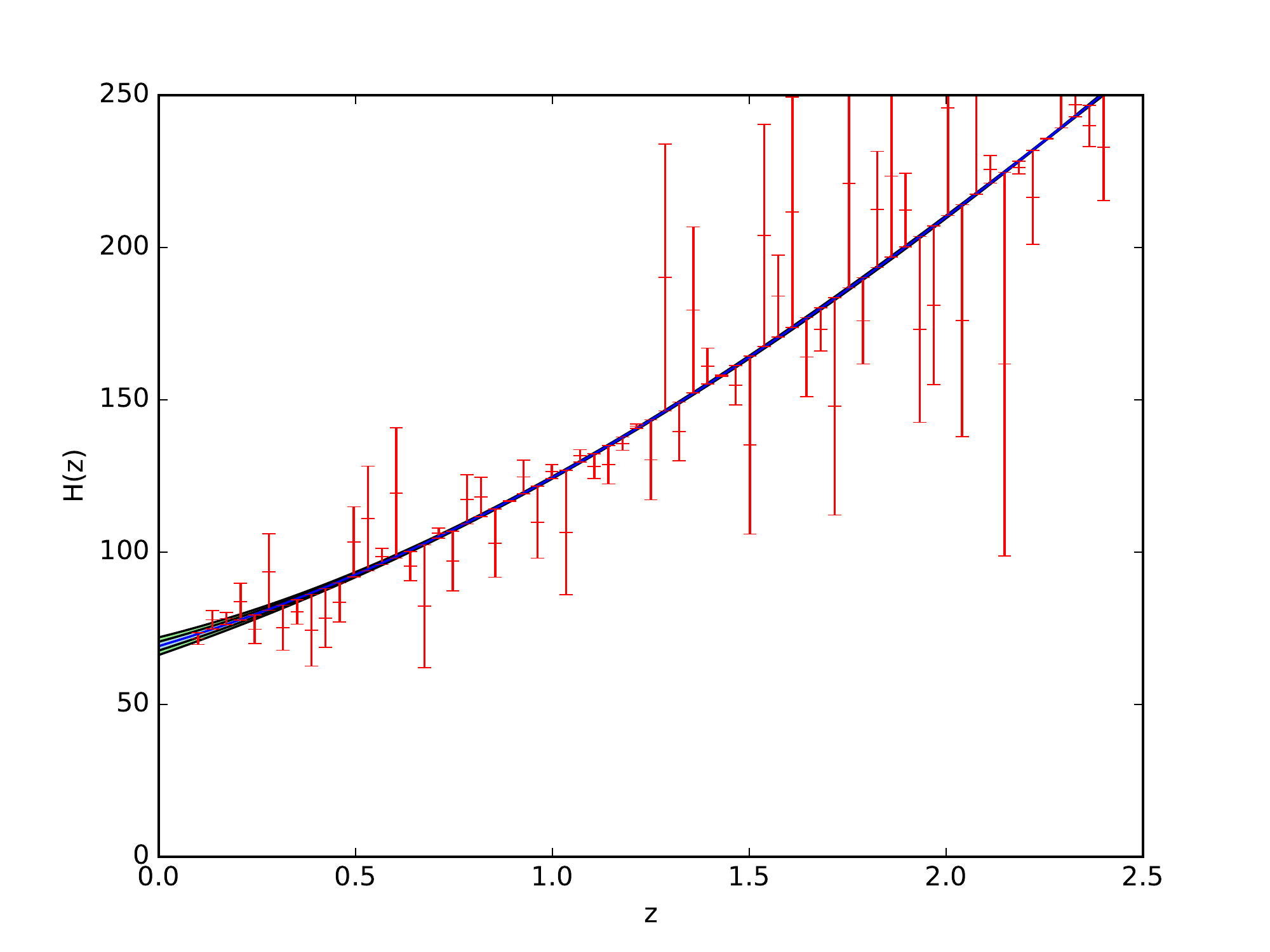}
\includegraphics[scale=0.3]{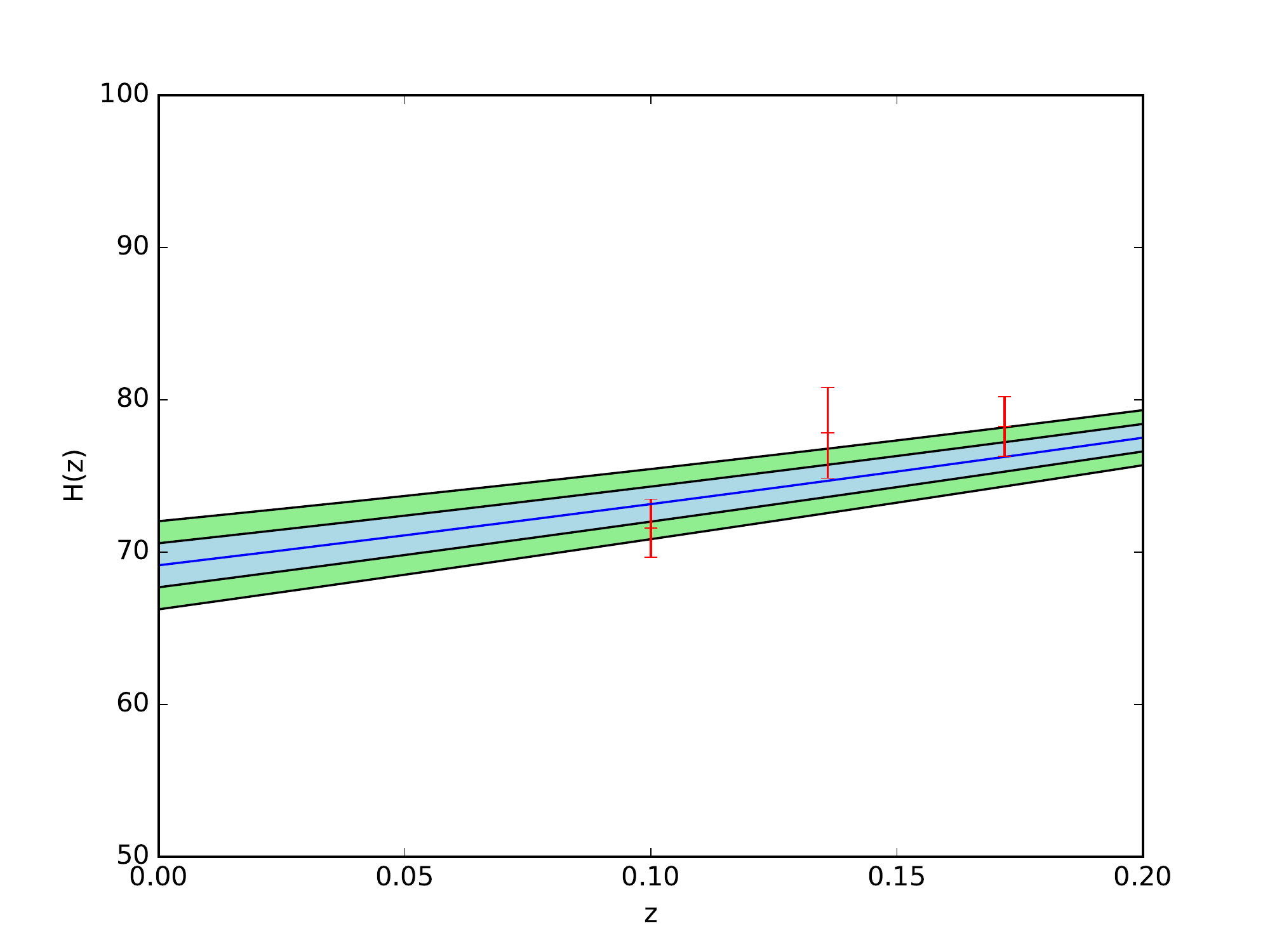}
\includegraphics[scale=0.3]{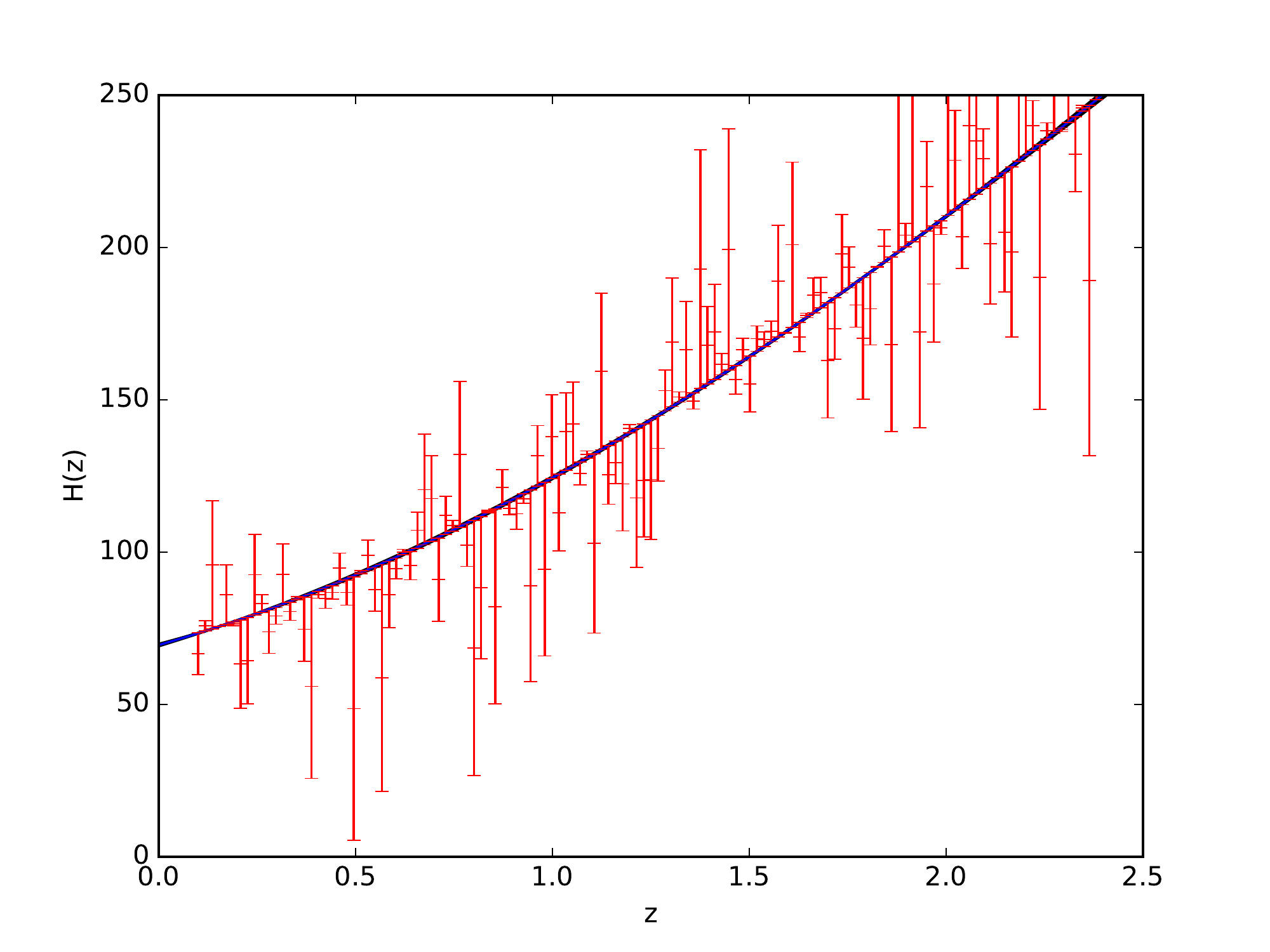}
\includegraphics[scale=0.3]{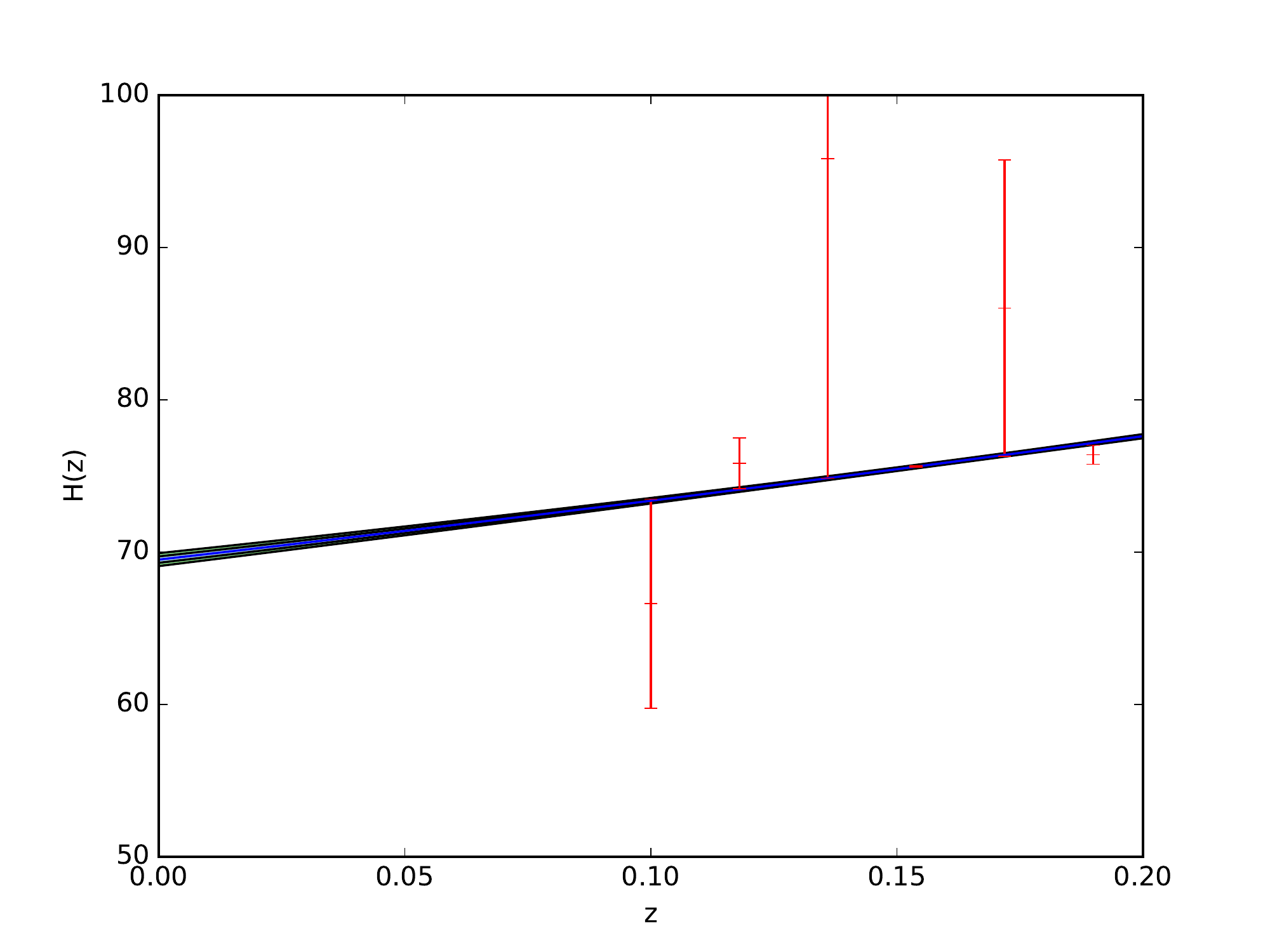}
\caption{From top to bottom, we exhibit the GP reconstructions of $H(z)$ using the simulated 64 and 128 $H(z)$ data points, respectively.}\label{f6}
\end{figure}

In this section, we utilize the extended 36 $H(z)$ measurements including the 19 measurements used in \cite{33} to implement our reconstruction, and the corresponding result is shown in Fig. \ref{f1}. When extrapolated to the redshift $z=0$, we find $H_0=69.21\pm3.72$ km s$^{-1}$ Mpc$^{-1}$. This indicates that the value of $H_0$ obtained by using the model-independent GP method is very consistent with the Planck 2015 and Riess et al. 2016 analysis at $1\sigma$ level. In Fig. \ref{f2}, using the larger sample of $H(z)$ data, one can also find that our result can resolve the $3.4\sigma$ tension more effectively than Busti et al. 2014 analysis. At the same time, our result is also compatible with Busti et al. 2014 analysis at $1\sigma$ confidence level.

As mentioned above, we would like to compare the GP reconstruction result with the standard parametric analysis. Adopting the usual $\chi^2$ statistics and using only the 36 $H(z)$ data points, we find $H_0=70.69\pm2.61$ km s$^{-1}$ Mpc$^{-1}$ for a flat $\Lambda$CDM model, $H_0=68.59\pm4.30$ km s$^{-1}$ Mpc$^{-1}$ for a flat $\omega$CDM model and $H_0=68.64\pm2.69$ km s$^{-1}$ Mpc$^{-1}$ for the decaying vacuum model. One can find that the value of $H_0$ for the $\Lambda$CDM case is only consistent with the local measurement, and the value of $H_0$ for the decaying vacuum case is only in agreement with the global measurement. Since the import of an extra parameter for the $\omega$CDM case, the value of $H_0$ has a larger error than the $\Lambda$CDM case, and is compatible with both the local and global measurements (see Fig. \ref{f3}).

To perform the systematic errors analysis of the GP method, as described in Ref. \cite{33}, we would like to consider three different effects on the GP method: the impact of outliers on determining $H_0$, different choices for the stellar population synthesis (SPS) model and the effects of different covariance functions on reconstructing $H_0$.

First of all, we consider the case of the existence of possible outliers. For the purpose to investigate the effects of high-redshift data points on determining $H_0$, in the upper panels of Fig. \ref{f4}, we exhibit the reconstruction by removing the whole data points with $z>1$. Different from the result in Ref. \cite{33}, we find that the result $H_0=68.22\pm3.36$ km s$^{-1}$ Mpc$^{-1}$ is still in agreement with the local and global measurements at $1\sigma$ level. This implies the low-redshift data is more reliable than the high-redshift data, and plays a primary role to determine $H_0$ in our extended $H(z)$ sample. In addition, one can also easily find that the behavior of $H(z)$ error blows up for the redshift range $z>1$, since we have removed the corresponding high-redshift data.  In order to study the impact of the data points with large errors on the reconstruction, in the lower panels of Fig. \ref{f4}, we implement the reconstruction by removing 6 data points with errors greater than 30 km s$^{-1}$ Mpc$^{-1}$. We find that the result $H_0=69.02\pm3.76$ km s$^{-1}$ Mpc$^{-1}$ is still compatible with the local and global measurements at $1\sigma$ level. Comparing it with the reconstruction result of the full sample $H_0=69.21\pm3.72$ km s$^{-1}$ Mpc$^{-1}$, one can easily conclude that the data points with large errors affect hardly the reconstruction.

\begin{table}[h!]
\begin{tabular}{ccccccc}
\hline
\hline
                          &$z$            & $H(z)$        & Ref.\\
\hline
                          &$0.070$        &$69\pm19.6$    & \cite{h1}                     \\
                          &$0.090$        &$69\pm12$      & \cite{h2}                    \\
                          &$0.120$        &$68.6\pm26.2$  & \cite{h1}                      \\
                          &$0.170$        &$83\pm8$       & \cite{20}                     \\
                          &$0.179$        &$75\pm4$       & \cite{h4}                     \\
                          &$0.199$        &$75\pm5$       & \cite{h4}                     \\
                          &$0.200$        &$72.9\pm29.6$    & \cite{h1}                     \\
                          &$0.270$        &$77\pm14$      & \cite{20}                     \\
                          &$0.240$        &$79.69\pm2.65$    & \cite{21}                     \\
                          &$0.280$        &$88.8\pm36.6$    & \cite{h1}                     \\
                          &$0.352$        &$83\pm14$    & \cite{h4}                     \\
                          &$0.3802$        &$83\pm13.5$    & \cite{h5}                     \\
                          &$0.400$        &$95\pm17$    & \cite{20}                     \\
                          &$0.4004$        &$77\pm10.2$    & \cite{h5}                     \\
                          &$0.4247$        &$87.1\pm11.2$    & \cite{h5}                     \\
                          &$0.430$        &$86.45\pm3.68$    & \cite{21}                     \\
                          &$0.4497$        &$92.8\pm12.9$    & \cite{h5}                     \\
                          &$0.4783$        &$80.9\pm9$    & \cite{h5}                     \\
                          &$0.480$        &$97\pm62$    & \cite{22}                     \\
                          &$0.570$        &$92.4\pm4.5$    & \cite{hh1}                     \\
                          &$0.593$        &$104\pm13$    & \cite{h4}                     \\
                          &$0.680$        &$92\pm8$    & \cite{h4}                         \\
                          &$0.781$        &$105\pm12$    & \cite{h4}                         \\
                          &$0.875$        &$125\pm17$    & \cite{h1}                         \\
                          &$0.880$        &$90\pm40$    & \cite{22}                         \\
                          &$0.900$        &$117\pm23$    & \cite{20}                         \\
                          &$1.037$        &$154\pm20$    & \cite{h4}                         \\
                          &$1.300$        &$168\pm17$    & \cite{20}                         \\
                          &$1.363$        &$160\pm33.6$    & \cite{h6}                         \\
                          &$1.430$        &$177\pm18$    & \cite{20}                         \\
                          &$1.530$        &$140\pm14$    & \cite{20}                         \\
                          &$1.750$        &$202\pm40$    & \cite{20}                         \\
                          &$1.965$        &$186.5\pm50.4$    & \cite{h6}                         \\
                          &$2.300$        &$224\pm8$    & \cite{h7}                         \\
                          &$2.340$        &$222\pm7$    & \cite{h8}                         \\
                          &$2.360$        &$226\pm8$    & \cite{h9}                         \\

\hline
\hline
\end{tabular}
\caption{The extended 36 H(z) measurements from different surveys using the differential age method and radial BAO method.}
\label{t1}
\end{table}

In the second place, we investigate the effect of different choices for the SPS model on determining $H_0$. In Ref. \cite{33}, the authors considered this effect for both Bruzual $\&$ Charlot (2003) (hereafter BC03) \cite{54} and Maraston $\&$ Str\"{o}mb\"{a}ck (2011) (hereafter M11) \cite{55} models by using 8 measurements from Ref. \cite{h4}. They found that the reconstruction results of $H_0$ depend obviously on the adopted SPS model. Moreover, the result for BC03 is very consistent with their full sample. It is noteworthy that these two models have substantial differences, for instance, the method utilized to estimate the integrated spectra, the treatment of the thermally pulsating asymptotical giant branch phase and the stellar evolutional models adopted to build the isochrones. Subsequently, we will use the extended 13 measurements by Moresco et al. \cite{h4,h5} to derive $H_0$ for both BC03 and M11 models. For BC03, we obtain $H_0=64.46\pm4.88$ km s$^{-1}$ Mpc$^{-1}$, which is very consistent with the result $H_0=64.4\pm4.9$ km s$^{-1}$ Mpc$^{-1}$ by Busti et al (see the upper panels of Fig. \ref{f5}). For M11, we find $H_0=69.83\pm4.98$ km s$^{-1}$ Mpc$^{-1}$, which is substantially different from the result $H_0=75.1\pm5.2$ km s$^{-1}$ Mpc$^{-1}$ by the same authors and can resolve well the $3.4\sigma$ tension at $1\sigma$ level between the local and global measurements (see the lower panels of Fig. \ref{f5}). This indicates that the M11 model is more sensitive to the newly added 5 data points, which lie in the redshift range [0.3702, 0.4783], than the BC03 model.

Another important element producing the systematic errors can be ascribed to the covariance functions in GaPP. In Ref. \cite{33}, the authors found that the reconstruction results of $H(z)$ depend obviously on the choice of covariance functions by using the 19 $H(z)$ measurements, and consequently affect the derived value of $H_0$. However, using the extended 36 H(z) measurements, we find that the concrete choice of covariance functions affects hardly the derived value of $H_0$. To show this better, we have performed the GP reconstruction of $H(z)$ using the 36 $H(z)$ measurements for the covariance functions Mat\'{e}rn $(5/2)$, Mat\'{e}rn $(7/2)$, Mat\'{e}rn $(9/2)$ and Cauchy, respectively. In Fig. \ref{f6}, one can easily conclude that the final reconstruction results of $H(z)$ are independent of the choice of covariance functions. This can be ascribed to the decreasingly statistical errors with the increasing sample size.

In addition, we still do not rule out the existence of new physics when analyzing the systematics of the GP method: because of some unknown physical mechanism, the derived values of $H_0$ from the reconstruction results have larger errors.

\section{The simulation}
To perform how the future data with what accuracy of error will affect the values of $H_0$ derived from our reconstruction results, we simulate 64 and 128 $H(z)$ data points, respectively, which lie in the redshift range [0.1, 2.4] and have the same data quality as the current 36 H(z) data points. In the current sample, we rule out 6 outliers with relatively large errors at low redshifts, and use the left 31 data points to estimate the errors of the simulated data (see Fig. \ref{f6}). Subsequently, we re-update the method of Ma et al. \cite{56} to generate the future $H(z)$ data by utilizing $H_{sim}(z)=H_{fid}(z)+N(0,\bar{\sigma}(z))$, where $H_{sim}(z)$, $H_{fid}(z)$ and $N(0,\bar{\sigma}(z))$ represent the simulated values of the Hubble parameter at redshift $z$, the fiducial values of the Hubble parameter at redshift $z$ and random numbers gaussianly distributed with mean zero and variance $\bar{\sigma}(z)$, respectively. We find that the uncertainties $\bar{\sigma}(z)$ are bounded by two straight lines: $\sigma_{+}(z)=20.48z+10.16$ and $\sigma_{-}(z)=2.07z+2.15$. If we believe the errors of future data are bounded by the two lines, we can take the mean line of the errors as $\sigma_{0}=11.26z+6.16$. Hence, the errors of the simulated data $\bar{\sigma}(z)$ obey a Gaussian distribution $N(\sigma_{0}(z),\eta(z))$, where $\eta(z)=[\sigma_{+}(z)-\sigma_{-}(z)]/4$ is chosen in order to assure the errors lie in the regions between $\sigma_{+}(z)$ and $\sigma_{-}(z)$ with $95.4\%$ probability.

In the upper panels of Fig. \ref{f7}, using the simulated 64 data points, we exhibit the GP reconstruction of $H(z)$ and obtain $H_0=69.08\pm1.74$ km s$^{-1}$ Mpc$^{-1}$. In the meanwhile, one can easily find that we have given the tighter constraint on $H(z)$ and reduced the uncertainty of $H_0$ from $5.4\%$ to $2.5\%$. In the lower panels of Fig. \ref{f7}, utilizing the simulated 128 data points, we get $H_0=70.54\pm0.28$ km s$^{-1}$ Mpc$^{-1}$ and the uncertainty of $H_0$ has been reduced to $0.4\%$. This indicates that for the same quality data, the more data points one simulates, the smaller the uncertainty of $H_0$ is. It is worth noticing that the value of $H_0$ derived from the reconstruction results of $H(z)$ depends strongly on the choice of $H_0$ in the fiducial model $H_{fid}(z)=H_0\sqrt{\Omega_{m0}(1+z)^3+(1-\Omega_{m0})}$, and for simplicity, we use $H_0=70$ km s$^{-1}$ Mpc$^{-1}$ here.

\section{discussions and conclusions}
The modern cosmological observations have provided more and more high-precision data. Recently, the new determination of the local value of the Hubble constant by Riess et al. 2016 has exhibited a strong tension with the global value derived from the CMB anisotropy data provided by the Planck satellite in the $\Lambda$CDM model. To alleviate or resolve this tension, we reapply the model-independent GP method to constrain $H_0$ by using the extended 36 $H(z)$ measurements, which includes the 19 measurements used by Busti et al. \cite{33}.      

Firstly, we obtain $H_0=69.21\pm3.72$ km s$^{-1}$ Mpc$^{-1}$, which is very consistent with the Planck 2015 and Riess et al. 2016 analysis at $1\sigma$ confidence level, and reduces the uncertainty from $6.5\%$ (Busti et al. 2014) to $5.4\%$. Subsequently, comparing our results with three parametric models, we find that the value of $H_0$ for the $\Lambda$CDM case is only consistent with the local measurement, and the value of $H_0$ for the decaying vacuum case is only compatible with the global measurement. Because of the import of an extra parameter for the $\omega$CDM case, the value of $H_0$ has a larger error than the $\Lambda$CDM case, and is in agreement with both the local and global measurements. In succession, we perform the systematic error analysis of the GP method and obtain the following conclusions: (i) removing the whole data points with $z>1$ to implement the reconstruction, we find that the low-redshift data is more reliable than the high-redshift data, and plays a primary role to determine $H_0$ in our extended $H(z)$ sample; (ii) removing 6 data points with errors greater than 30 km s$^{-1}$ Mpc$^{-1}$, we conclude that the data points with large errors affect hardly the reconstruction; (iii) using the extended 13 measurements, we find that the M11 model is more sensitive to the newly added 5 measurements, which lie in the redshift range [0.3702, 0.4783], than the BC03 model; (iv) different from the results of Ref. \cite{33} by using 19 $H(z)$ measurements, we find that the final reconstruction results of $H(z)$ are independent of the choice of covariance functions, which can be ascribed to the decreasingly statistical errors with the increasing sample size. Moreover, we can not rule out the existence of new physics when analyzing the systematics of the GP method: due to some unknown physical mechanism, the values of $H_0$ derived from the reconstruction results have larger errors. Furthermore, we utilize the simulated data to investigate how the future data with what accuracy of error will affect the values of $H_0$ derived from our reconstruction results. Without loss of generality, assuming $H_0=70$ km s$^{-1}$ Mpc$^{-1}$ in the fiducial model $H_{fid}(z)$, we find that for the same quality data, the more data points one simulates, the smaller the uncertainty of $H_0$ is. To be more precise, the uncertainty of $H_0$ has been reduced from
$5.4\%$ to $0.4\%$. This can be ascribed to the simulations which not only increase the number of the same quality data points but also improve the data quality at high redshifts.
 
In the future, with more and more high-precision data, we expect to constrain the value of $H_0$ better by using other statical methods or developing new cosmological models, in order to alleviate or resolve the strong tension between the local and global measurements. 

\section{acknowledgements}
This study is partly supported by the National Science Foundation of China. We are thankful to Professors S. D. Odintsov and Bharat Ratra for beneficial communications on cosmology. The author Deng Wang warmly thanks Prof. Jing-Ling Chen for talks on quantum information, Qi-Xiang Zou for helpful discussions, Pu-Yuan Gao for programming.

\end{document}